\documentclass[sn-nature]{sn-jnl}

\usepackage{graphicx}%
\usepackage{multirow}%
\usepackage{amsmath,amssymb,amsfonts}%
\usepackage{amsthm}%
\usepackage{mathrsfs}%
\usepackage[title]{appendix}%
\usepackage{xcolor}%
\usepackage{textcomp}%
\usepackage{manyfoot}%
\usepackage{booktabs}%
\usepackage{algorithm}%
\usepackage{algorithmicx}%
\usepackage{algpseudocode}%
\usepackage{listings}%

\usepackage{lineno}
\usepackage{geometry}
\usepackage{tabularx}
\usepackage{siunitx}
\DeclareSIUnit \year {yr}
\DeclareSIUnit \erg  {erg}
\DeclareSIUnit \parsec  {pc}
\DeclareSIUnit \gauss  {G}

\newcommand\editcyan[1]{\textcolor{black}{#1}}

\theoremstyle{thmstyleone}%
\theoremstyle{thmstyletwo}%

\theoremstyle{thmstylethree}%

\begin{document}

\title[Article Title]{\centering Suppression of pair beam instabilities in a \\ laboratory analogue of blazar pair cascades}

\author*[1,2]{\fnm{Charles D.} \sur{Arrowsmith}}\email{charles.arrowsmith@physics.ox.ac.uk}
\author[1]{\fnm{Francesco} \sur{Miniati}}
\author[3]{\fnm{Pablo J.} \sur{Bilbao}}
\author[4,5]{\fnm{Pascal} \sur{Simon}}
\author[1]{\fnm{Archie F. A.} \sur{Bott}}
\author[4]{\fnm{Stephane} \sur{Burger}}
\author[6]{\fnm{Hui} \sur{Chen}}
\author[3]{\fnm{Filipe D.} \sur{Cruz}}
\author[7]{\fnm{Tristan} \sur{Davenne}}
\author[1]{\fnm{Anthony} \sur{Dyson}}
\author[4]{\fnm{Ilias} \sur{Efthymiopoulos}}
\author[2]{\fnm{Dustin H.} \sur{Froula}}
\author[4]{\fnm{Alice} \sur{Goillot}}
\author[8,9]{\fnm{Jon T.} \sur{Gudmundsson}}
\author[2]{\fnm{Dan} \sur{Haberberger}}
\author[1,7]{\fnm{Jack W. D.} \sur{Halliday}}
\author[1,10]{\fnm{Tom} \sur{Hodge}}
\author[1]{\fnm{Brian T.} \sur{Huffman}}
\author[1]{\fnm{Sam} \sur{Iaquinta}}
\author[10]{\fnm{G.} \sur{Marshall}}
\author[11]{\fnm{Brian} \sur{Reville}}
\author[1]{\fnm{Subir} \sur{Sarkar}}
\author[1]{\fnm{Alexander A.} \sur{Schekochihin}}
\author[3]{\fnm{Luis O.} \sur{Silva}}
\author[6]{\fnm{Raspberry} \sur{Simpson}}
\author[1,4,12]{\fnm{Vasiliki} \sur{Stergiou}}
\author[7]{\fnm{Raoul M. G. M.} \sur{Trines}}
\author[11]{\fnm{Thibault} \sur{Vieu}}
\author[4]{\fnm{Nikolaos} \sur{Charitonidis}}
\author[7,13]{\fnm{Robert} \sur{Bingham}}
\author[1]{\fnm{Gianluca} \sur{Gregori}}

\affil[1]{\orgdiv{Department of Physics}, \orgname{University of Oxford}, \orgaddress{\street{Parks Road}, \city{Oxford}, \postcode{OX1 3PU}, \country{UK}}}

\affil[2]{\orgname{University of Rochester Laboratory for Laser Energetics}, \orgaddress{\city{Rochester}, \state{NY},\postcode{14623}, \country{USA}}}

\affil[3]{\orgdiv{GoLP/Instituto de Plasmas e Fusão Nuclear, Instituto Superior Técnico}, \orgname{Universidade de Lisboa}, \orgaddress{\postcode{1049-001}, \city{Lisboa}, \country{Portugal}}}

\affil[4]{\orgname{European Organization for Nuclear Research (CERN)}, \orgaddress{\street{CH-1211 Geneva 23}, \country{Switzerland}}}

\affil[5]{\orgname{GSI Helmholtzzentrum für Schwerionenforschung GmbH}, \orgaddress{\street{Planckstraße 1
64291 Darmstadt}, \country{Germany}}}

\affil[6]{\orgname{Lawrence Livermore National Laboratory}, \orgaddress{\street{7000 East Ave}, \city{Livermore}, \state{California}, \postcode{94550}, \country{USA}}}

\affil[7]{\orgname{STFC Rutherford Appleton Laboratory}, \orgaddress{\street{Chilton}, \city{Didcot}, \postcode{OX11 0QX}, \country{UK}}}

\affil[8]{\orgdiv{Science Institute}, \orgname{University of Iceland}, \orgaddress{\street{Dunhaga 3}, \postcode{IS-107}, \city{Reykjavik},  \country{Iceland}}}

\affil[9]{\orgdiv{Division of Space and Plasma Physics, School of Electrical Engineering and Computer Science}, \orgname{KTH Royal Institute of Technology}, \orgaddress{ \postcode{SE-100 44}, \city{Stockholm},  \country{Sweden}}}

\affil[10]{\orgname{AWE}, \orgaddress{\street{Aldermaston}, \city{Reading}, \state{Berkshire}, \postcode{RG7  4PR}, \country{UK}}}

\affil[11]{\orgname{Max-Planck-Institut für Kernphysik}, \orgaddress{\street{Saupfercheckweg 1}, \postcode{D-69117}, \city{Heidelberg}, \country{Germany}}}

\affil[12]{\orgdiv{School of Applied Mathematics and Physical Sciences}, \orgname{National Technical University of Athens}, \orgaddress{\postcode{Athens 157 72}, \country{Greece}}}

\affil[13]{\orgdiv{Department of Physics}, \orgname{University of Strathclyde}, \orgaddress{\city{Glasgow}, \postcode{G4 0NG}, \country{UK}}}

\maketitle
\centerline{\large Dated: \today }

\newpage

\section*{Abstract}

{\bf The generation of dense electron-positron pair beams in the laboratory can enable direct tests of theoretical models of $\gamma$-ray bursts and active galactic nuclei. We have successfully achieved this using ultra-relativistic protons accelerated by the Super Proton Synchrotron at CERN. In the first application of this experimental platform, the stability of the pair beam is studied as it propagates through a metre-length plasma, analogous to TeV $\gamma$-ray induced pair cascades in the intergalactic medium. It has been argued that pair beam instabilities disrupt the cascade, thus accounting for the observed lack of reprocessed GeV emission from TeV blazars. If true this would remove the need for a moderate strength intergalactic magnetic field to explain the observations. We find that the pair beam instability is suppressed if the beam is not perfectly collimated or monochromatic, hence the lower limit to the intergalactic magnetic field inferred from $\gamma$-ray observations of blazars is robust.
}

\section*{Significance statement}
In this work, a dense beam of electron-positron pairs is produced using protons accelerated by the Super Proton Synchrotron at CERN. The beam is propagated through an ambient plasma, analogous to pair cascades produced as blazar jets propagate through the intergalactic medium (IGM). It has been proposed that plasma instabilities disrupt these pair cascades, explaining the lack of secondary $\gamma$-rays observed from blazars. However, we find that under non-ideal conditions likely to be relevant in the blazar context, pair beam instabilities are strongly suppressed and it is unlikely they play a significant role. This experimental study supports the hypothesis that the IGM contains a magnetic field of unknown origin that may well be a relic of the early Universe.

\section*{Main}


TeV-blazars are a class of active galactic nuclei (AGN) with relativistic jets pointing towards Earth that emit $\gamma$-rays with a hard spectrum extending to TeV energies \cite{2013ApJ...764..119S}. As the $\gamma$-rays propagate through the intergalactic medium (IGM), often through cosmic voids where matter exists as a tenuous, collisionless plasma, they are expected to scatter with extragalactic background light and trigger electromagnetic cascades of electron-positron (e$^{\pm}$) pairs. Subsequently, inverse-Compton scattering of the pairs with cosmic microwave background (CMB) photons is expected to produce a reprocessed spectrum of GeV energy $\gamma$-rays, but this is at odds with astronomical observations accumulated over more than a decade which have set stringent limits on the expected GeV $\gamma$-rays~\cite{aharonian2006low}. The leading hypothesis is that e$^{\pm}$ pairs are deflected by intergalactic magnetic fields (IGMF)~\cite{neronov2010evidence}, spreading the GeV emission into diffuse halos that are not resolved by $\gamma$-ray telescopes \cite{archambault2017search,ackermann2018search} (but may be detectable by the Cherenkov Telescope Array \cite{2021JCAP...02..048A}). However, the required strength and coherence length of the IGMF is sufficiently large that an astrophysical origin is unlikely; it may well be a relic of the early Universe \cite{2018PhRvL.121b1301M,2021RPPh...84g4901V,hosking2023cosmic}. An alternative suggestion is that a substantial fraction of the pairs' energy is dissipated via electromagnetic beam-plasma instabilities~\cite{weibel1959spontaneously,fried1959mechanism,fainberg1970nonlinear,1971JETP...32.1134R} before the pairs inverse-Compton scatter with CMB photons~\cite{broderick2012cosmological,schlickeiser2012plasma}. Such instabilities lead to exponential amplification of electromagnetic fields via the unstable separation of electrons and positrons into current filaments. In extreme scenarios, they can lead to significant dissipation of the bulk kinetic energy and the onset of collisionless shocks. Whether these indeed play a role in the blazar-induced pair cascade, and similar situations involving streaming e$^{\pm}$ pairs such as $\gamma$-ray bursts, depends on the initial linear stage of the instability (characterized by a linear growth rate), and the subsequent quasi-linear evolution which determines the transition to the saturation stage. When non-idealized conditions are considered, for example finite divergence and energy spread, plasma kinetic theory suggests that beam-plasma instabilities may be strongly suppressed. 

For relativistic pair beams with a low density compared with the ambient plasma, the fastest-growing modes of electromagnetic beam instability are oriented obliquely to the beam direction \cite{shaisultanov2011stream}, with a characteristic scale comparable to the plasma skin depth, $\lambda_{\mathrm{s}}=c/\omega_{\mathrm{p}}$, where $\omega_{\rm p} = (4\pi n_{\rm p} e^2/m_{\rm e})^{1/2}$ is the plasma frequency, $n_{\rm p}$ is the ambient plasma electron density, $e$ is the elementary charge and $m_{\mathrm{e}}$ is the electron rest mass. The theoretical growth rate for a pair beam with a small angular spread $\Delta\theta = \Delta p_{\perp}/p_{\parallel}$ is given by~\cite{fainberg1970nonlinear}:
\begin{equation}
\Gamma = \frac{\sqrt{3}}{2^{4/3}}\,\omega_{\rm p}\bigg(\frac{n_{\pm}}{n_{\rm p}\gamma_{\pm}}\bigg)^{1/3}\left[\frac{k_{\perp}^{2/3}}{k^{2/3}}-\frac{3k_{\perp}^2(\Delta\theta)^2}{8k_{\parallel}^2}\bigg(\frac{2n_{\rm p}\gamma_{\pm}}{n_{\pm}}\bigg)^{2/3}\right],
\label{eq:oblique}
\end{equation}
where $k_{\parallel}$ and $k_{\perp}$ are the parallel and perpendicular components of the wavevector relative to the beam's propagation axis, $n_{\pm}$ is the pair number density and $\gamma_{\pm}$ is the relativistic Lorentz factor. 
The factor in square brackets accounts for transverse thermal motion of the beam, and tends to unity for a sufficiently collimated beam given $\Delta\theta\ll(n_{\pm}/n_{\rm p}\gamma_{\pm})^{1/3}$, with the fastest-growing modes oriented transversely to beam propagation. However, if the transverse momentum spread is large enough, i.e. $\Delta\theta\gtrsim(n_{\pm}/n_{\rm p}\gamma_{\pm})^{1/3}$, particle trajectories traverse the characteristic scale of unstable modes before current fluctuations can be amplified. The effect is to stabilize small-scale transverse modes, reduce the fastest growth rate, and tilt the direction of the fastest-growing mode towards the longitudinal direction. This is important in astrophysical contexts, where relativistic pair beams may be neither perfectly collimated nor monoenergetic. The purely transverse filamentation instability growth rate (scaling as $\Gamma\sim\omega_{\rm p}\sqrt{n_{\rm \pm}/n_{\rm p}\gamma_{\rm \pm}}$~\cite{fried1959mechanism}) only applies under very strict conditions, namely when the beam and return current are perfectly symmetric with the same density, temperature, and drift energies~\cite{bret2010multidimensional}. If the angular spread is sufficiently large ($\Delta\theta>\Gamma\lambda_{\rm s}/c$) then the purely transverse filamentation modes are completely stabilized~\cite{silva2003interpenetrating}, whereas oblique unstable modes that lead to beam filamentation and magnetic field growth at an angle with respect to the beam distribution can still exist with a suppressed growth rate. In the kinetic regime, with very large transverse spreads the growth rate scales as $\Gamma\sim \omega_{\rm p}(n_{\rm \pm}/n_{\rm p} \gamma_{\rm \pm})/(\Delta\theta)^2$ \cite{breizman1971quasilinear,bret2010exact}, even more severely suppressed than predicted by Eq.~\ref{eq:oblique}.

In order for pair beam instabilities to provide a plausible explanation for the observed lack of secondary GeV $\gamma$-rays from TeV blazars, the instability must lead to the generation of sufficiently large electromagnetic fields for a substantial fraction of the pairs’ energy to be dissipated, or for pairs to be sufficiently deflected by self-generated fields that the secondary GeV emission is dispersed. 
This scenario has recently been discussed when considering whether the 400 GeV photon detected by Fermi from the “BOAT” GRB~221009A~\cite{axelsson2025grb} may be generated from a TeV electromagnetic cascade in the IGM, a possibility consistent with the observation by LHAASO of $\gamma$-rays with energy up to 13~TeV from the direction of GRB~221009A~\cite{lhaaso2023very}. However, the short 33~ks time delay of the 400 GeV photon implies intergalactic magnetic fields significantly weaker than the lower bounds derived from blazar observations~\cite{ackermann2018search}; this tension would be resolved if pair beam instabilities play a significant role. 
Nevertheless, it has been argued that suppression of the instability growth rates due to the large momentum spread of the cascade prevents their significant development before the pairs inverse-Compton scatter with CMB photons~\cite{miniati2013relaxation,sironi2014relativistic,perry2021role}. This conjecture is difficult to verify using simulations due to their limited spatial and temporal range, nor has the theory been validated experimentally due to the challenge of producing pair beams with the necessary high density and quasi-neutrality~\cite{chen2023perspectives}. But experimental studies now become possible due to our recent breakthrough demonstration that relativistic e$^{\pm}$ pair beams can be efficiently produced using ultra-relativistic proton beams accelerated by the Super Proton Synchrotron (SPS) at CERN~\cite{arrowsmith2023laboratory}. Here, we investigate the stability of e$^{\pm}$ pair beams as they propagate through a metre-length ambient plasma to test whether the beam-plasma instabilities are indeed suppressed under non-idealized conditions, analogous to astrophysical situations involving e$^{\pm}$ pairs streaming through the IGM. We discuss the implications for blazar pair cascades in cosmic voids and comment on the robustness of the lower limit to the IGMF inferred from observations of blazar $\gamma$-ray spectra. 

\begin{figure}[t!]
\centering
\includegraphics[width=1\textwidth]{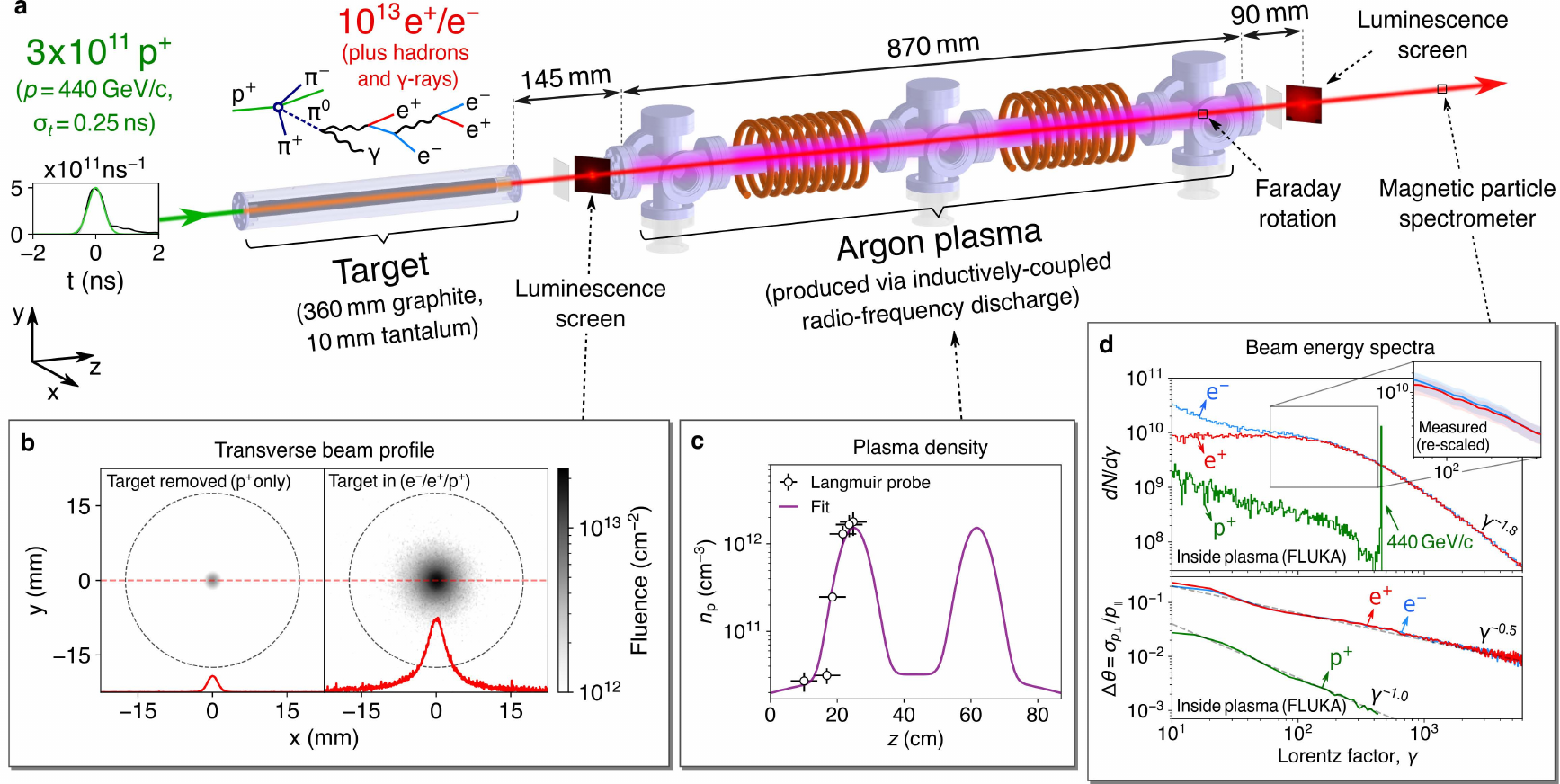}
\caption{{\bf Experimental setup.} 
(a) Protons with \SI{440}{\giga\electronvolt}/$c$ momentum are extracted from the SPS accelerator with temporal profile measured using an integrating current transformer (inset). The protons irradiate a solid target (\SI{360}{\milli\metre} graphite plus \SI{10}{\milli\metre} tantalum) with a maximum fluence exceeding $3\times10^{11}$ protons in a single bunch of duration \SI{0.25}{\nano\second} (1-$\sigma$) and transverse size $\sigma_r = \SI{1}{\milli\metre}$. A secondary beam is generated via hadronic and electromagnetic cascades, containing a dominating fluence of electron-positron pairs ($N_{\mathrm{e}^{\pm}}>10^{13}$), plus hadrons, $\gamma$-rays and other secondaries. (b) Measurements of the transverse beam profile using $70$~mm~$\times\, 50$~mm~$\times \, 0.25$~mm chromium-doped ceramic (Chromox) luminescence screens, viewed by a digital camera at a standoff distance of $\SI{3.8}{\metre}$. The spatial resolution is limited to $\SI{100}{\micro\metre}$ by the Chromox transluscence. (c) Downstream of the target, the beam passes though a metre-length argon plasma produced by an inductively-coupled radio-frequency discharge. The longitudinal plasma density profile is measured using a Langmuir probe and confirmed during the experiment using optical emission spectroscopy. Pair beam filamentation due to beam-plasma instability is measured using a luminescence screen placed downstream of the plasma (camera positioned at a standoff distance \SI{3.9}{\metre}),  and a Faraday rotation diagnostic measures the growth of magnetic fields inside the plasma. (d) The electron and positron energy spectra are measured using a magnetic particle spectrometer~\cite{arrowsmith2023laboratory}, shown here re-scaled according to the size of the collecting aperture to compare directly with FLUKA simulations of the spectra inside the plasma. 
}
\label{fig:Experimental_Setup}
\end{figure}

The experimental setup is shown in Figure~\ref{fig:Experimental_Setup}. More than $3\times10^{11}$ protons are extracted from the SPS and delivered to the HiRadMat (High-Radiation to Materials) facility~\cite{efthymiopoulos2011hiradmat} with momentum \SI{440}{\giga\electronvolt}/c in a single LHC-type bunch (transverse size $\sigma_r=\SI{1}{\milli\metre}$ and duration $\sigma_t=\SI{0.25}{\nano\second}$). The proton beam irradiates a custom-designed solid target consisting of a graphite rod with a tantalum converter. Hadronic interactions of the protons with carbon nuclei generates a copious number of neutral pions ($\pi^0$), which decay to produce a highly collimated beam of GeV-energy $\gamma$-rays. Inside the tantalum, the $\gamma$-rays trigger electromagnetic cascades of relativistic electron-positron pairs, leading to a secondary quasi-neutral beam containing over $10^{13}$ relativistic e$^{\pm}$ pairs, along with a much smaller number of protons and other secondary products~\cite{arrowsmith2021generating}. Characterization of the secondary beam is provided by Monte-Carlo simulations performed using FLUKA~\cite{ahdida2022new,battistoni2015overview,vlachoudis2009flair}, a standard code capable of accurately describing hadronic and electromagnetic cascades in the target. The simulations are validated against luminescence screen measurements of the transverse beam profile and magnetic spectrometer measurements of the e$^{\pm}$ energy spectra~\cite{arrowsmith2023laboratory}. The pairs exhibit a multi-power-law spectrum in momentum, $dN_{\pm}/dE\propto E^{-m}$, with spectral index $m\approx1-2$. The beam then propagates through an inductively coupled argon discharge plasma, with the plasma density and temperature measured prior to the experiment using a Langmuir probe, and confirmed non-invasively during the experiment using optical emission spectroscopy (details provided in the Supplementary Information). The plasma density is characterized by two identical bumps reaching a peak density of $n_{\mathrm{p}}=(1.8\pm0.1)\times\SI{e12}{\per\centi\metre\cubed}$ when the discharge vessel is filled with argon to a pressure $p_{\mathrm{g}}=\SI{4}{\pascal}$ and the inductive coils are supplied with 1 kW of radio-frequency power, corresponding to an absorbed power $P_{\mathrm{abs}}=240\pm\SI{10}{\watt}$~\cite{arrowsmith2023inductively}. Importantly, the physical size of the pair beam exceeds the expected size of current filaments (on the order of the plasma skin depth). 
Similar to many astrophysical situations, the ambient plasma is relativistically cold ($k_{\mathrm{B}}T_{\mathrm{e}}\sim\si{\electronvolt}$), and the rates of electron-neutral collisions ($\nu_{\mathrm{en}}$) and electron-ion collisions ($\nu_{\mathrm{ei}}$) are much smaller than the plasma frequency (calculations in the Supplementary Information). 

Since a collisionless kinetic description of the plasma applies to both the experiment and the astrophysical case, similarity scaling arguments may be applied~\cite{ryutov1999similarity}. \editcyan{In this regard we take care to consider the possible effects of the residual proton beam which accompanies the pair beam; though these protons are much smaller in number than the e$^{\pm}$ pairs, they remain localized to the central axis providing an additional current (and azimuthal magnetic field). We show in the Supplementary Information how the beam-plasma instability differs between our case and for pure, quasi-neutral e$^{\pm}$ jets. The residual current seeds the initial growth of the instability, but no other dynamical effects are observed and the instability grows at a comparable rate.}
Two main diagnostics are employed in the experiment: (i) a time-resolved, magneto-optic Faraday rotation probe positioned near the end of the plasma to measure magnetic fields generated by pair beam instability, and (ii) a chromium-doped alumina luminescence screen positioned downstream of the plasma to detect modulations in the transverse beam profile arising from the formation of electron/positron filaments (see Methods).

\subsection*{Experimental results}

\begin{figure}[t!]
\centering
\includegraphics[width=0.5\textwidth]{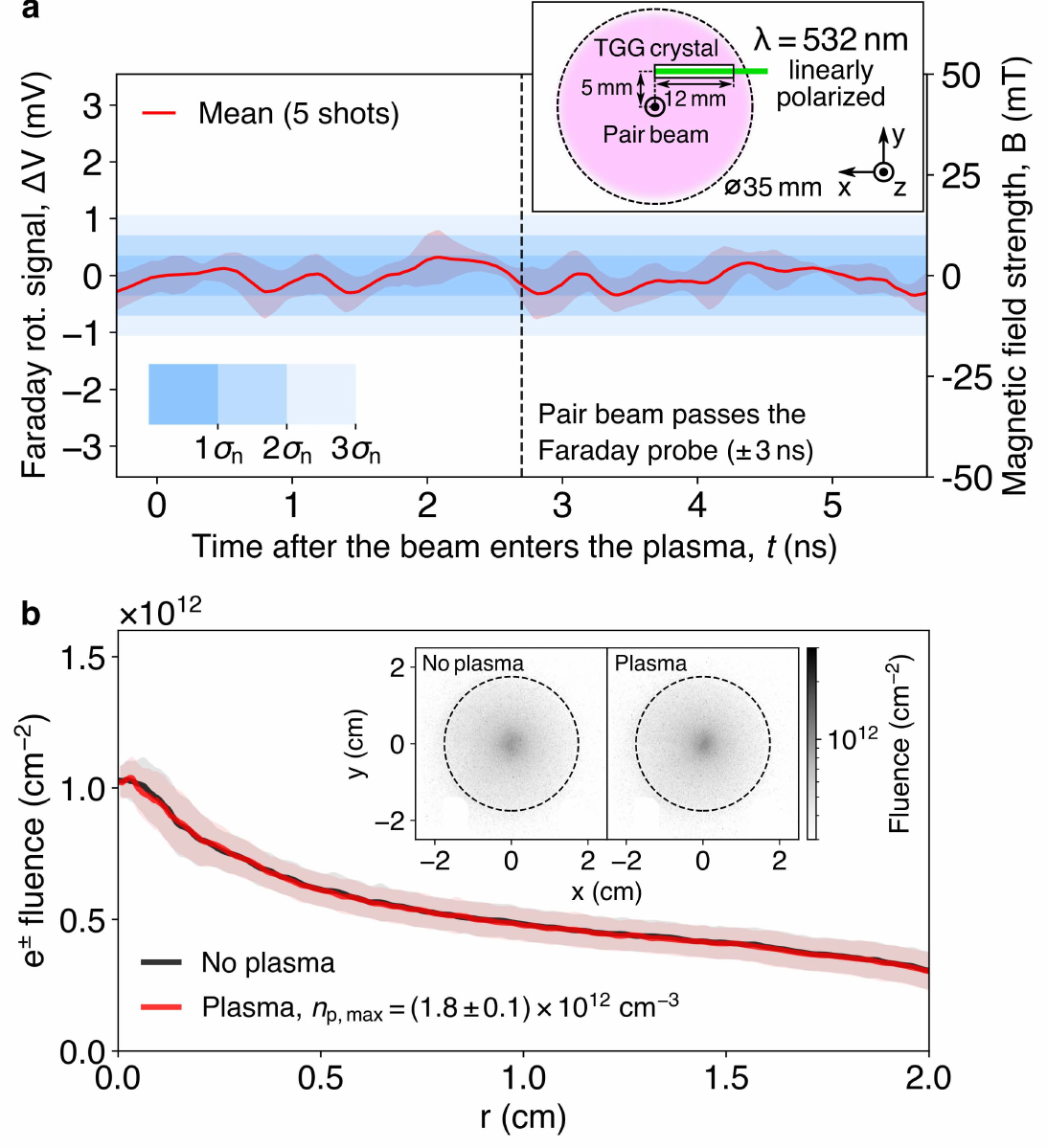}
\caption{{\bf Magnetic field and beam profile measurements.} (a) Magnetic fields are measured at the end of the plasma using a time-resolved Faraday rotation technique. A linearly polarized laser beam ($\lambda=\SI{532}{\nano\metre}$) is passed twice through a magneto-optic crystal (TGG, \SI{12}{\milli\metre} length, \SI{2}{\milli\metre} diameter), suspended in the plasma by a ceramic re-entrant tube (orientation shown in the inset), before passing through a second polarizing filter offset by $\SI{45}{\degree}$ from the initial polarization. The change in laser intensity measured by a fast photodiode ($\Delta V = V-V_0/2$) is plotted as a function of time, with the pair beam expected to pass the probe $\SI{2.7}{\nano\second}$ after the beam enters the plasma (the signal is shown a few ns before and afterwards to account for the uncertainty in the exact timing). The mean signal from five shots is plotted, with corresponding standard deviation represented by the red shaded region. The blue shaded regions show the standard deviation of the intrinsic electronic noise. (b)~The transverse beam profile is measured using a Chromox luminescence screen positioned $\SI{90}{\milli\metre}$ downstream of the plasma discharge. The residual primary proton beam is subtracted from the images (see Methods), leaving the fluence of electron-positron pairs (plus additional secondaries). The radial lineouts are shown when the plasma is present ($p_{\mathrm{g}} = \SI{4}{\pascal}$, $P_{\mathrm{abs}}=240\pm\SI{10}{\watt}$) and when there is no plasma ($p_{\mathrm{g}} = \SI{0.5}{\pascal}$, $P_{\mathrm{abs}}=\SI{0}{\watt}$), with the image data shown in the inset. The shaded regions represent the standard deviation of the lineout pixel counts combined with the uncertainty in the absolute calibration. 
}
\label{fig:data}
\end{figure}

The experimental results of both diagnostics are summarized in Figure~\ref{fig:data}. The inset of Figure~\ref{fig:data}a shows the orientation of the magneto-optic terbium gallium garnet (TGG) crystal used to perform the Faraday rotation measurement, suspended in the plasma \SI{81}{\centi\metre} downstream of the beam entry. A linearly polarized laser beam ($\lambda=\SI{532}{\nano\metre}$) is passed twice through the TGG crystal, reflecting at the rear surface. Magnetic fields oriented along the length of the crystal will cause the polarization of the laser beam to rotate, which is detected by a time-resolved measurement of the intensity of the laser after passing through a second polarizing filter oriented at $45^{\circ}$ to the initial polarization. The light is collected by a fast photodiode ($t_{10-90\%}=\SI{0.44}{\nano\second}$) and the measured intensity is given by $V = V_0\cos^2(\langle B\rangle \mathcal{V}L + 45^{\circ})$, where $\langle B\rangle= \frac{2}{L}\int \mathbf{B} \cdot \mathbf{\hat{x}} \,dx$ is the mean component of the magnetic field along the propagation axis of the laser beam, $\mathcal{V}=217\pm\SI{15}{\radian\per\tesla\per\metre}$ is the measured Verdet constant, $L=\SI{12}{\milli\metre}$ is the length of the crystal, and $V_0=\SI{14}{\milli\volt}$ is the intensity measured when the two polarizing filters are aligned and $\langle B\rangle=0$. The prefactor of 2 accounts for the double pass of the crystal. Magnetic fields can be measured on the timescales of the beam duration ($t_{10-90\%}=\SI{0.42}{\nano\second}$) with a single-shot sensitivity limited to $B_{\rm sens}=\SI{5}{\milli\tesla}$ by the intrinsic electronic noise (full characterization provided in the Supplementary Information). Since the timing is consistent between shots, the signal-to-noise ratio can be improved by combining signals from multiple shots. Five measurements of the pair beam are obtained and the mean intensity change ($\Delta V$) and standard deviation are plotted in Figure~\ref{fig:data}a, showing that the maximum deviation of the mean signal corresponds to $\langle B\rangle=\SI{5}{\milli\tesla}$, weak enough that a precise measurement remains limited by the electronic noise floor. We infer an upper limit of the measured field of $\langle B\rangle_{\mathrm{exp}}\leq\SI{5}{\milli\tesla}$, given that a value consistently larger than $\SI{5}{\milli\tesla}$ in five repeated measurements leads to a statistically significant detectable signal (at the 2.2-$\sigma$ level). An upper bound of the instability growth rate is estimated assuming that if the growth rate is sufficiently small then the magnetic field grows exponentially from the azimuthally oriented seed field, $B_0$, produced by the net current of the residual proton beam propagating on-axis: $\langle\Gamma_{\mathrm{exp}}\rangle=t_{\mathrm{prop}}^{-1}\log\left[\langle B\rangle_{\mathrm{exp}}/B_0\right]\leq\SI{0.7}{\per\nano\second}$, where $t_{\mathrm{prop}}=\SI{2.7}{\nano\second}$ is the propagation time of the pair beam through plasma, and $B_0 = 0.78\pm\SI{0.13}{\milli\tesla}$ is calculated from the precise net current distribution obtained from a FLUKA simulation (as detailed in the Supplementary Information). 
For comparison, the beam instability growth rate calculated from linear kinetic theory (using Eq.~\ref{eq:oblique}) assuming a perfectly collimated beam gives a linear growth rate $\Gamma = \SI{2.0}{\per\nano\second}$, significantly higher than observed in the experiment. 

It is possible that the Faraday probe measurement may underestimate the strength of the magnetic field if large-amplitude variations are present on scales smaller than the length of the Verdet crystal, however this would be accompanied by observations of strong separation of current in the measurements of the downstream transverse beam profile (Figure~\ref{fig:data}b), for example, clear structures of the kind observed in particle-in-cell simulations of collimated pair beams presented in the next section. Instead, no difference is observed between the beam profiles with and without the plasma present. Low-contrast current separation can be obscured due to the equal sensitivity of the screen to electrons and positrons, but the development of high-contrast, millimetre-scale filaments can be ruled out, and a slower-than-expected growth rate remains the most plausible explanation. We attribute the reduced growth rate to suppression due to the finite thermal spread of the pairs, and we show in the following section that particle-in-cell simulations predict that the instability of collimated, monoenergetic pairs would instead produce much stronger, measurable magnetic fields. 
A reduction of the pair beam's density as it diverges may also cause the growth rate to be reduced, but theoretical calculations indicate that this cannot explain the full extent of the observed suppression. Additional measurements were performed with lower peak plasma densities and with the probe oriented to point in a radial direction to the beam path (presented in the Supplementary Information). In all cases, similarly weak magnetic fields were observed.

\subsection*{Particle-in-cell simulations}

\begin{figure}[b!]
\centering
\includegraphics[width=1\textwidth]{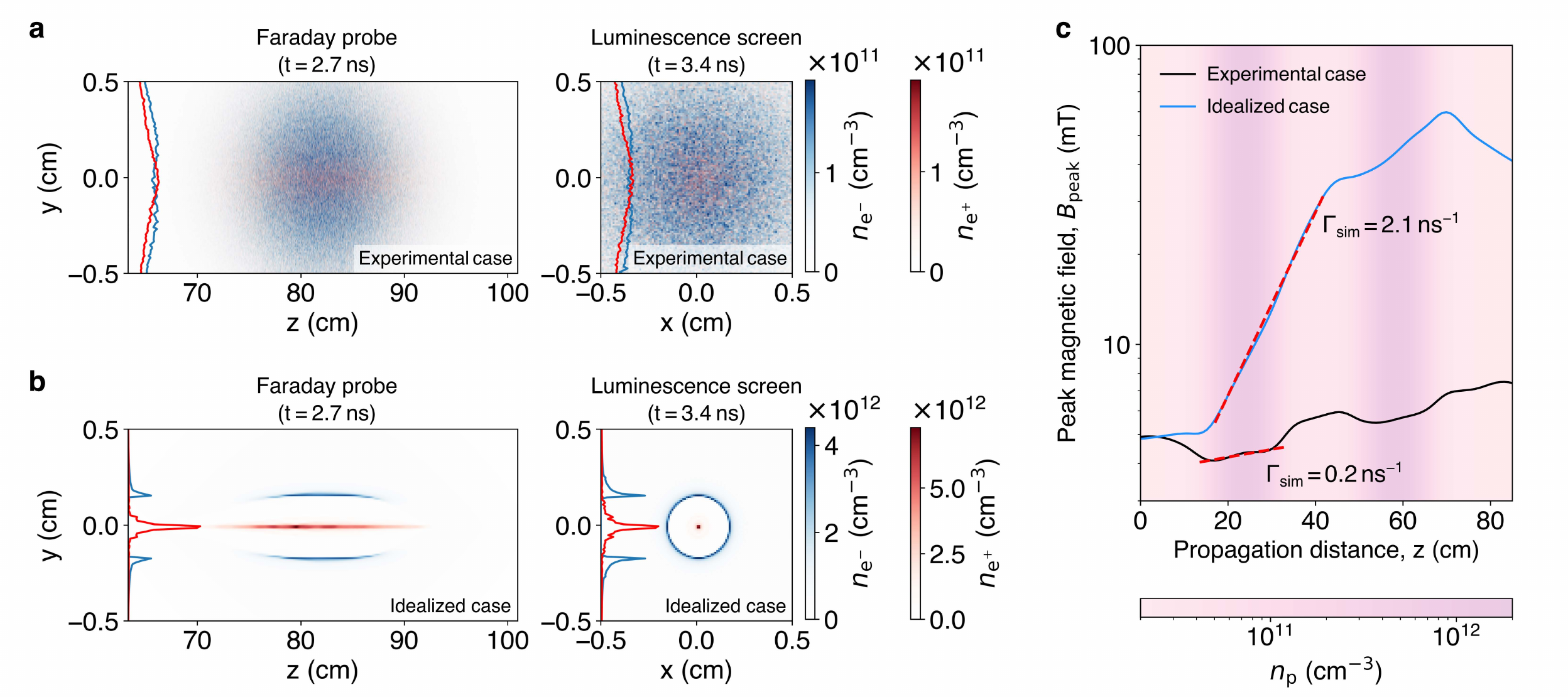}
\caption{{\bf Three-dimensional particle-in-cell simulations.} 
3D simulations of the beam-plasma interaction are performed using the particle-in-cell code OSIRIS~\cite{fonseca2002osiris} for two cases: (a) with conditions closely resembling the experimental beam and ambient plasma (labelled `experimental'), and (b) an idealized case with a collimated, monoenergetic ($\gamma_{\pm}=10^3$) pairs (labelled `idealized'). In (a) and (b), the left panels show a central slice of the electron and positron densities in the longitudinal plane ($y$-$z$) at the time when the beam passes the Faraday probe ($t=\SI{2.7}{\nano\second}$ after entering the plasma), whilst the right panels show a central slice of the transverse plane ($x$-$y$) when the beam passes the downstream luminescence screen ($t=\SI{3.4}{\nano\second}$). (c) The peak magnetic field is plotted as a function of propagation distance through the plasma, with the background shading showing the ambient plasma density. The maximum growth rate of the peak magnetic field is obtained by the shown fits (red-dashed). The anti-correlation of magnetic field and plasma density is an effect of the varying level of return current.
}
\label{fig:3DPIC}
\end{figure}

The effects of finite thermal spread of the e$^{\pm}$ pairs are explored further by performing three-dimensional (3D3V) particle-in-cell (PIC) simulations using the fully relativistic, massively parallel PIC code OSIRIS \cite{fonseca2002osiris}. In the simulations, a moving window follows an electron-positron-proton beam at the speed of light to model conditions closely resembling the experimental pair beam and the ambient plasma (see Methods). Given the initial charge and current distribution of the beam species (electron, positrons and protons), as determined from the FLUKA simulations, the corresponding self-consistent fields are determined by solving Poisson's equations in the beams' proper frame, and then Lorentz boosting these fields back fo the laboratory frame, ensuring the fields are self-consistently coupled to the beams. The idealized case of a perfectly collimated monoenergetic pair beam is considered for comparison. In simulations performed without the residual primary protons, a quasi-neutral pair beam is observed to filament with fluctuations that grow from thermal noise without regard for the central axis (as shown in additional simulations presented in the Supplementary Information). When they are included, the protons co-propagate on-axis providing a seed magnetic field for the instability, which causes the positron density to increase on-axis and the azimuthal magnetic field to be amplified exponentially with the same growth rate. Otherwise, the protons do not play a dynamical role during the experiment timescale because of their much larger inertia. \editcyan{When identical simulations are performed without a background plasma, instabilities are no longer observed.}

The results are shown in Figure~\ref{fig:3DPIC}. In both cases, current separation of electrons and positrons is observed on the scale of the plasma skin depth. 
In the case where the e$^{\pm}$ pairs are collimated and monoenergetic, the filaments become completely separated and emerge on much smaller spatial scales. 
The peak magnetic field is vastly increased ($B_{\rm max}=\SI{60}{\milli\tesla}$) with a peak growth rate $\Gamma_{\rm sim}=\SI{2.1}{\per\nano\second}$, well-matched by the theoretical prediction \editcyan{for a cold pair beam ($\Gamma = \SI{2.0}{\per\nano\second}$, Eq.~\ref{eq:oblique})}.
By contrast, when the finite thermal spread of pairs is accounted for, the small-scale modes are stabilized, and the peak magnetic field and growth rate are significantly reduced ($B_{\rm max}=\SI{7}{\milli\tesla}$, $\Gamma_{\rm sim}=\SI{0.2}{\per\nano\second}$). The simulated magnetic fields provide a prediction of the Faraday rotation measurement of $\langle B\rangle_{\mathrm{sim}}=\SI{1.7}{\milli\tesla}$, leading to an estimated average growth rate $\langle\Gamma_{\mathrm{sim}}\rangle=\SI{0.3}{\per\nano\second}$, consistent with the experimentally obtained bound $\langle\Gamma_{\mathrm{exp}}\rangle\leq\SI{0.7}{\per\nano\second}$.

\subsection*{Discussion}

To assess whether beam-plasma instabilities are important for blazar-induced pair beams propagating through cosmic voids, a maximum growth rate of the instability is estimated by scaling the experimental upper bound on the growth rate according to Eq.~\ref{eq:oblique}. The pair density and energy spectrum of a typical blazar pair cascade are estimated using a Monte-Carlo model, described in Elyiv~et~al.~\cite{elyiv2009gamma} and used by e.g. Neronov~\&~Vovk~\cite{neronov2010evidence} and Miniati~\&~Elyiv~\cite{miniati2013relaxation} (modeled parameters given in Table~\ref{tab:parameters}). The blazar spectral emission is assumed to be a power-law distribution, $\mathrm{d}N_{\gamma}/\mathrm{d}E_{\gamma}\propto E_{\gamma}^{-1.8}$ in the range $10^3\leq E_{\gamma}/m_{\rm e}c^2 \leq 10^8$, with equivalent isotropic luminosity \SI{e45}{\erg\per\second}, using a model for the extragalactic background light described by Aharonian~\cite{2001ICRC...27I.250A}. The pair density obtained is model dependent, but $n_{\pm,\mathrm{blz}}\sim10^{-23}\,(1+z)^{9.5}\,\si{\per\centi\metre\cubed}$~\cite{broderick2012cosmological,miniati2013relaxation} is considered to be reasonable, where $z$ is the redshift. This redshift dependence is expected to be valid for TeV blazars with $z\lesssim1$~\cite{broderick2012cosmological}. The mean Lorentz factor of pairs is $\langle\gamma_{\pm}\rangle\sim10^5$, whilst pairs that inverse-Compton scatter with CMB photons to produce GeV emission have a much higher Lorentz factor ($\gamma_{\pm}\sim10^7$). 
The density of free electrons in the void is estimated by $n_{\mathrm{p,IGM}}=\Omega_{\mathrm{b}}f_{\mathrm{IGM}}\rho_{\mathrm c}/ m_{\mathrm{p}}$, where $\Omega_{\mathrm{b}}$ is the cosmological baryon density parameter, $f_{\mathrm{IGM}}$ is the relative fraction of baryons in the IGM, $m_{\mathrm{p}}$ is the proton rest mass, and $\rho_{\mathrm c}=3H_0^2/8\pi G$ is the critical mass density of the Universe, where $H_0$ is the Hubble constant and $G$ is the gravitational constant. The IGM is assumed to be fully ionized~\cite{meiksin2009physics}. The fraction of baryons in the IGM, $f_{\mathrm{IGM}}$, is constrained by measurements of the dispersion measure of extragalactic fast radio bursts to be $f_{\mathrm{IGM}}\approx0.4-0.8$~\cite{li2019cosmology,lemos2023cosmological} (more recently, $f_{\mathrm{IGM}} = 0.76^{+0.10}_{-0.11}$~\cite{connor2025gas}). A void density $n_{\mathrm{p,IGM}}=2\times10^{-7}\,(1+z)^{3}\,\si{\per\centi\metre\cubed}$ is obtained by assuming $f_{\mathrm{IGM}}=0.8$ and using standard cosmological parameters from Planck~\cite{aghanim2020planck}: $\Omega_{\mathrm{b}}h^2=0.02237\pm0.00015$, where $h = H_0/\SI{100}{\kilo\metre\per\second\per\mega\parsec}$. 

\begin{figure}[h!]
\centering
\includegraphics[width=0.6\textwidth]{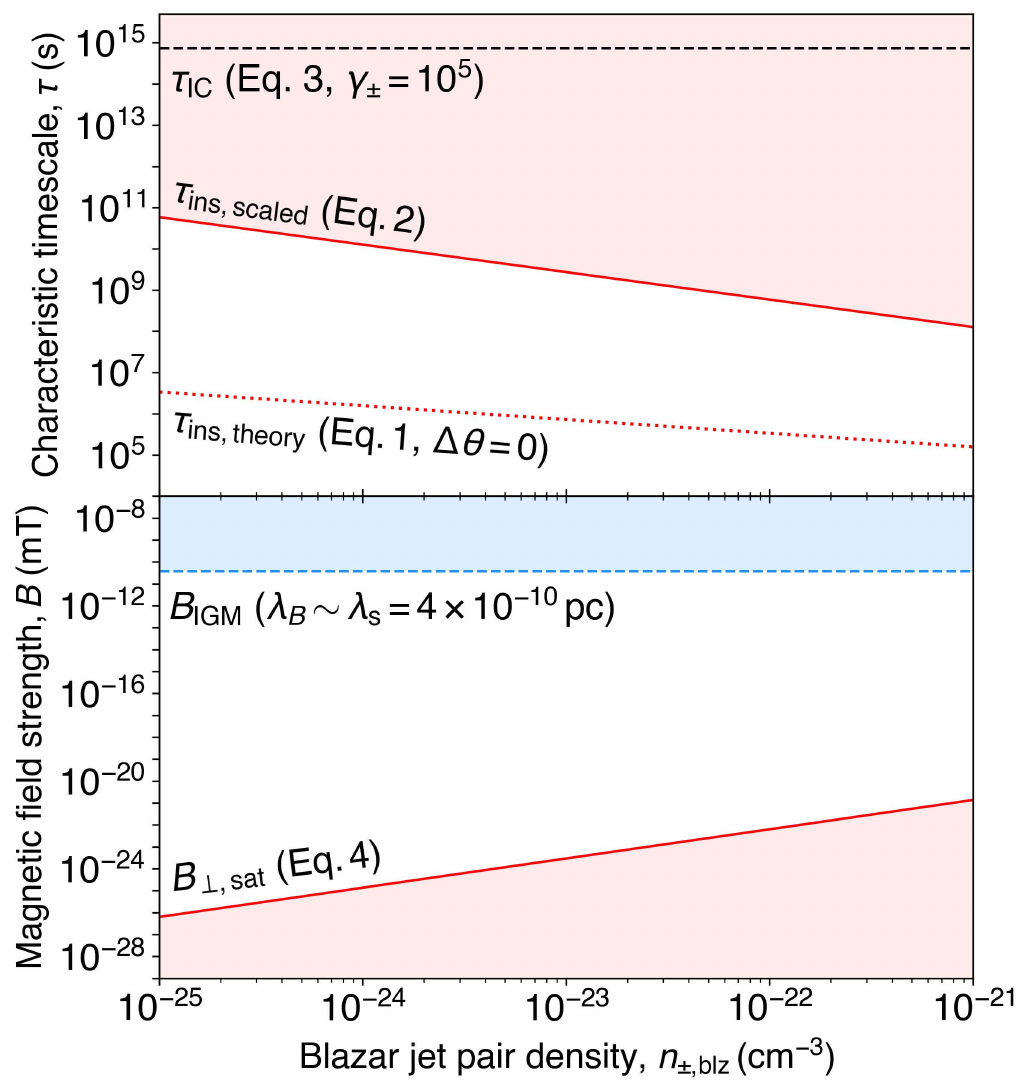}
\caption{{\bf Characteristic timescales and magnetic field strengths relevant to blazar-induced pair cascades in cosmic voids.} Upper panel: The experimentally obtained bound on the growth rate of electromagnetic pair beam instability ($\Gamma_{\mathrm{exp}}\leq\SI{0.7}{\per\nano\second}$) is scaled using Eq.~\ref{eq:scaling} to obtain a lower bound on the linear growth rate for blazar-induced pair beams propagating through cosmic void (red solid). The growth timescale is much larger than theoretical predictions assuming a collimated, monoenergetic pair beam (Eq.~\ref{eq:oblique}, $\Delta\theta=0$, red dotted), but short compared with the inverse-Compton cooling time of pairs with CMB photons (Eq.~\ref{eq:IC}, $\gamma_{\pm}=10^5$, black-dashed). 
Lower panel: The maximum strength of the magnetic field at instability saturation is inferred assuming the experimentally obtained bound on the growth rate (Eq.~\ref{eq:Bsat}, red solid). The field strengths are incompatible with lower bounds inferred from blazar spectral measurements for a coherence length comparable to the ambient skin depth (blue-dashed). It is assumed in all cases that the redshift is $z\lesssim1$.
}
\label{fig:scaling}
\end{figure}

Under these conditions, electromagnetic beam instability is strongly suppressed ($\Delta\theta\gg(n_{\pm}/n_{\rm p}\gamma_{\pm})^{1/3}$), and we can scale the linear growth rate according to Eq.~\ref{eq:oblique} in the limit of large $\Delta\theta$ ($\Gamma = \sqrt{2/3}\,\omega_{\rm p}(n_{\pm}/2n_{\rm p}\gamma_{\pm})^{2/3}/(\Delta\theta)$):
\begin{equation}
\Gamma_{\rm blz}\,\mbox{[s$^{-1}$]}
\leq 4\times 10^{-10}\,\bigg(\frac{\Gamma_{\rm exp}}{\SI{0.7}{\per\nano\second}}\bigg).
\label{eq:scaling}
\end{equation}
This estimate likely overestimates the growth rate of the oblique filamentation instability, comparing to predictions for the kinetic regime where the equivalent scaling gives $\Gamma_{\rm blz}\leq \SI{2e-13}{\per\second}$.
Assuming the more conservative bound, the linear growth timescale ($\tau_{\rm ins} \equiv 1/\Gamma_{\rm blz}$) is plotted in Figure~\ref{fig:scaling} alongside estimates of the fastest growth rate for a collimated, monoenergetic blazar-jet pair cascade and the inverse-Compton cooling time of pairs with CMB photons:
\begin{equation}
\tau_{\rm IC}\,\mbox{[s]} 
=3.8\times10^{13}~\bigg(\frac{E_{\rm e}}{\SI{1}{\tera\electronvolt}}\bigg)^{-1}(1+z)^{-4}.
\label{eq:IC}
\end{equation}
For blazar-jet pair densities in the range $n_{\pm,\mathrm{blz}}=10^{-25}-\SI{e-21}{\per\centi\metre\cubed}$, we find that the scaled suppressed growth rate may still be fast compared with the inverse-Compton cooling rate. However, the saturated strength of magnetic fields is limited by the particle-trapping condition~\cite{davidson1972nonlinear}. That is, magnetic field generation is expected to saturate when the frequency at which particles ``bounce'' between magnetic filaments is comparable to the instability growth rate prior to saturation, i.e., when $\omega_{B}\sim \Gamma$, where $\omega_{B}\sim v/\sqrt{\lambda_{B}r_{\rm g}}$, $v$ is the particle velocity, $\lambda_{ B}$ is the length scale of magnetic filaments and $r_{\rm g}$ is the gyroradius. Assuming $\lambda_{B}$ is comparable to the skin depth of the ambient plasma~\cite{bret2010multidimensional,iwamoto2023kinetic}:
\begin{equation}
    B_{\perp,\rm{sat}}\,\mbox{[mT]}\lesssim 4\times10^{-24}\,\bigg(\frac{\Gamma_{\rm blz}}{\SI{4e-10}{\per\second}}\bigg)^2.
\label{eq:Bsat}
\end{equation}
We compare this field strength in Figure~\ref{fig:scaling} with lower limits on the IGMF imposed by the lack of observed GeV cascade emission in Fermi/LAT and MAGIC telescope data: 
$B_{\rm{IGM}}\,\mbox{[mT]}\geq 4\times10^{-11}\,(\lambda_{B}/\SI{4e-10}{\parsec})^{-1/2}$. We find that even if many $e$-folding lengths of instability can develop, the generated magnetic fields are much too small to explain the required angular spreading. After saturation of electromagnetic modes, longitudinal electrostatic oscillations may continue to grow at the slower quasi-linear rate~\cite{breizman1971quasilinear}, but the velocity spread of pairs is much too large to allow significant coupling of the pair beam's kinetic energy into large-amplitude, resonantly driven plasma waves (i.e. $k\cdot\Delta v_{\pm}\gg\Gamma$). 
We can therefore rule out that beam-plasma instabilities plausibly affect TeV blazar pair cascades; hence the inferred lower bound on the intergalactic magnetic field strength is robust.

\begin{table}[h!]
\caption{{\bf Pair cascade and plasma parameters in the experiment and blazar jets in cosmic voids.}}\label{tab:parameters}%
\begin{tabularx}{\textwidth}{@{}lll@{}}
\toprule
Parameter & Experiment & Typical blazar jet (luminosity \SI{e45}{\erg\per\second}) \\ &&at a distance \SI{300}{\mega\parsec} \\
\midrule
\textbf{Plasma}&&\\
Ambient plasma density, $n_{\rm p}\,$(cm$^{-3}$) & $10^{12}$   & $2\times10^{-7}$ \\
Skin depth, $c/\omega_{\rm p}$ (m) & $5\times 10^{-3}$& $10^7$ \\
Collisionality, $\nu_{\rm e}/\omega_{\rm p}$    & $10^{-3}$   & $10^{-13}$   \\
\midrule
\textbf{Pair cascade}&&\\
Pair density, $n_{\pm}\,$(cm$^{-3}$)  & $5\times10^{10}$   & $10^{-23}$  \\
Mean Lorentz factor, $\langle\gamma_{\pm}\rangle$    & $10^{3}$   & $10^{5}$  \\
Transverse momentum spread, $\Delta \theta $  & $0.025$  & $10^{-4}$   \\
\botrule
\end{tabularx}
\end{table}

\clearpage

\section*{Materials and methods}

\subsection*{Electron-positron pair production target}

The target used to produce the secondary beam consists of a \SI{360}{\milli\metre} length cylinder of isostatic graphite (SGL Carbon R6650, \SI{1.84}{\gram\per\cubic\centi\metre}) and a \SI{10}{\milli\metre} thickness tantalum converter, both with a diameter of \SI{20}{\milli\metre}. The graphite and tantalum are encased inside a \SI{400}{\milli\metre} length, \SI{50}{\milli\metre} diameter cylinder of high-strength T9 aluminium alloy that acts as both a confinement vessel and a heat sink. The tantalum is press-fit to ensure maximal thermal contact. \SI{2}{\milli\metre} thickness expanded graphite pieces (SGL Carbon Sigraflex, \SI{1}{\gram\per\cubic\centi\metre}) separate the target components to allow thermal expansion and reduce contact stresses during irradiation, while \SI{2}{\milli\metre} thickness Sigradur G glassy carbon beam windows are clamped onto either end of the target by aluminium flanges with Viton O-rings to seal the target materials hermetically. Radiative and convective cooling via the outer surface of the target housing leads to cooling of the target to room temperature within a few seconds following the beam impact, while the beam-induced maximum strain of the tantalum remains in all cases well below its plastic deformation limit, i.e. the target is not destroyed when irradiated by the proton beam and can be reused for many (potentially thousands of) shots.

\subsection*{Inductively coupled argon plasma discharge}

The plasma discharge is composed of a vacuum chamber constructed from three six-way port crosses (non-ferrous, 304L-grade stainless steel), separated by two sections of glass tube (\SI{15}{\centi\metre} length), terminated at each end by a \SI{4}{\milli\metre}-thick glassy carbon beam window. The total length of the discharge region is $L=\SI{87}{\centi\metre}$, with an inner diameter $d = \SI{3.5}{\centi\metre}$. Inductive coils wrap around the sections of glass tube (\SI{1}{\centi\metre} diameter copper pipe, with 8.5~turns, and coil-winding inner diameter $\SI{7.5}{\centi\metre}$). The coils and surrounding metallic segments of the chamber are connected to a zero ground potential. The vacuum pump, argon gas line and Faraday probe are attached at the port crosses. Before operation, the chamber is evacuated to a base pressure $p_{\rm g,base} = \SI{5}{\milli\pascal}$, before filling with argon gas (purity 99.999\%) to a pressure $p_{\rm g} =\SI{4}{\pascal}$. Radio-frequency power is supplied to the coils at a frequency $f=\SI{13.56}{\mega\hertz}$ using a commercially available 1 kW radio-frequency power generator (Advanced Energy CESAR 1310) via an impedance-matching network (Advanced Energy Navio). The power absorbed by the plasma corresponds to approximately 25\% of the supplied power. A plasma is produced by inductive coupling inside the coils with plasma density exceeding $n_{\rm p}\gtrsim\SI{e12}{\per\centi\metre\cubed}$, whilst a lower density plasma ($n_{\rm p}\sim10^{10}-\SI{e11}{\per\centi\metre\cubed}$) is sustained between the coils by electrostatic/capacitive coupling. The discharge is ignited several seconds before the beam's arrival and is deactivated several seconds after the beam has passed. Further details and extensive plasma characterization are provided in Ref.~\cite{arrowsmith2023inductively}.

\subsection*{Time-resolved magneto-optic Faraday rotation probe}

A continuous-wave diode laser (Z-Laser ZM18, $\lambda = \SI{532}{\nano\metre}$, $P = \SI{40}{\milli\watt}$) is transported to the Faraday probe's location via optical fibre and linearly polarized using a nanoparticle linear film polarizer, chosen for its high extinction ratio (10,000:1), high transmission ratio (73\%) at $\lambda=\SI{532}{\nano\metre}$ and high damage threshold. The laser beam is split using a 50:50 beamsplitter and the transmitted beam is used as a reference beam to monitor changes in the intensity of the probe laser beam, whilst the reflected beam is directed transversely to the particle-beam axis through a \SI{12}{\milli\metre} length, \SI{2}{\milli\metre} diameter terbium-gallium-garnet (TGG) magneto-optic crystal suspended in the plasma at the end of a ceramic re-entrant tube (inner length \SI{70}{\milli\metre}, outer diameter \SI{5}{\milli\metre}). The crystal is positioned \SI{81}{\centi\metre} into the plasma at a closest distance of \SI{5}{\milli\metre} to the particle-beam axis (orientation shown in Figure~\ref{fig:data}). The crystal has an anti-reflective coating on the front surface and a highly reflective coating on the rear surface to reflect the probe beam so that it makes a double pass of the crystal, thereby doubling the Faraday rotation for a given magnetic field. The reflected probe beam passes through a second linear polarizer oriented at a $45^{\circ}$ angle to the first and is coupled into a \SI{7}{\metre} length optical fibre (silica core, glass clad, step index with 0.22\,NA, \o\SI{200}{\micro\metre} core). The light is collected by a \SI{2}{\giga\hertz} bandwidth photodiode (Thorlabs DET025AFC/M) and a shielded coaxial cable connects the photodiode to a \SI{3}{\giga\hertz} oscilloscope (LeCroy WavePro 7300A). 
A schematic of the design is provided in the Supplementary Information along with measurements of the Verdet constant, intrinsic noise and instrument response function, measuring instrument sensitivity to magnetic fields of magnitude $ B_{\rm sens}=\SI{5}{\milli\tesla}$ and a time resolution of \SI{0.44}{\nano\second} ($10-90\%$ rise time). The Verdet constant of the crystal is measured again at the end of the beam time to confirm that the instrument's sensitivity has not degraded due to radiation damage.

\subsection*{Chromium-doped alumina luminescence screens}

Chromium-doped alumina-ceramic luminescence screens (Chromox, Al$_2$O$_3$:~99.5\%, Cr$_2$O$_3$:~0.5\%) are used to measure the transverse beam profile before and after the beam passes through the plasma. When energy is deposited in the screen by ionizing particles and radiation, luminescence light is emitted isotropically, strongest at wavelengths $\lambda_1 = \SI{691}{\nano\metre}$, and $\lambda_2=\SI{694}{\nano\metre}$ with decay times $3-\SI{6}{\milli\second}$. Screens are oriented at 45$^{\circ}$ to the beam path and viewed directly by a digital camera (Basler acA1920-40gm GigE camera with Sony IMX249 CMOS sensor and Canon EF $75-\SI{300}{\milli\metre}$ f/4-5.6 III lens) at a standoff distance \SI{4}{\metre} with an exposure time \SI{24}{\milli\second}. It is reasonably assumed that the vast majority of particles incident on the screens are relativistic and deposit approximately the same amount of energy in a minimum-ionizing fashion. The spatial resolution is limited due to screen translucence to $\SI{100}{\micro\metre}$ (attenuation length, $\mu=\SI{0.8}{\per\milli\metre}$), capable of resolving the filament formation in the pair beam since the size of unstable modes is expected to be several mm. The screen downstream of the plasma is positioned $d_{\rm scr}=\SI{90}{\milli\metre}$ from the glassy carbon beam window, and a blocker foil (\SI{50}{\micro\metre} aluminium) is placed before the screen to minimize stray optical light. In Figure~\ref{fig:data}, the residual protons in the transverse beam profile are subtracted from the raw image by fitting the proton peak to a 2D Gaussian with initial parameters $\sigma_x=\sigma_y=\SI{1}{\milli\metre}$ and integrated intensity corresponding to the expected number of residual primary protons that do not significantly scatter inelastically with the target, a fraction given by $N_{\mathrm{res}}/N_{\mathrm{inc}} =\exp(-L_{\mathrm{C}}/\lambda_{\mathrm{C}})\exp(-L_{\mathrm{Ta}}/\lambda_{\mathrm{Ta}})=0.42$, where $L_{\mathrm{C}} = \SI{360}{\milli\metre}$ and $L_{\mathrm{Ta}} = \SI{10}{\milli\metre}$ are the lengths of graphite and tantalum, and $\lambda_{\mathrm{C}}= \SI{466}{\milli\metre}$ and $\lambda_{\mathrm{Ta}}= \SI{115}{\milli\metre}$ are their corresponding nuclear interaction lengths. The calculation of $N_{\mathrm{res}}$ is in agreement with FLUKA Monte-Carlo simulations. Coulomb scattering of e$^{\pm}$ pairs in the glassy carbon beam window is only expected to significantly affect the pairs with a much lower energy than the mean, with a point source diverging to a size $\sigma_{\mathrm{spread}}\approx \SI{0.3}{\milli\metre}\,(\gamma_{\pm}/\langle \gamma_{\pm} \rangle)^{-1}$~\cite{lynch1991approximations}. 

\subsection*{Particle-in-cell (PIC) simulations}
Three-dimensional (3D3V) PIC simulations were performed using the OSIRIS code at the exascale LUMI supercomputer (Finland). Simulations use a moving window travelling at $c$ along the $z$-direction that follows relativistic electrons, positrons and protons in the secondary beam as they propagate through the ambient plasma. The electron-positron-proton beam is initialised before entrance of the plasma, centred at $z= \SI{-20}{\centi\metre}$ and $x=y=0$. The density and momentum phase-space distributions are accurately modelled by fitting analytical forms to the distributions at the entrance of the plasma cell, after the glassy carbon window, obtained from a FLUKA simulation (as described in the Supplementary Information). The longitudinal density profile of the plasma is chosen to match closely the measured electron density profile of the plasma discharge (plotted alongside Langmuir probe data in the inset of Figure~\ref{fig:Experimental_Setup}): double peaked with maximum plasma density $n_0=\SI{1.78e12}{\centi\metre^{-3}}$ (functional form provided in the Supplementary Information). All quantities in the simulations are normalized to the peak plasma density $n_0$ (associated plasma period $\omega_{\rm pe}^{-1}=\SI{13.29}{\pico \second}$, and plasma skin-depth $c/\omega_{\rm pe}=\SI{3.98}{\milli\metre}$). The moving window has absorbing boundary conditions and dimensions $L_x\times L_y\times L_z = \SI{3.5}{\centi\metre} \times \SI{3.5}{\centi\metre}\times \SI{40}{\centi\metre}$, discretised into $879\times879\times10050$ cells. This yields a spatial resolution $\Delta x = 0.01\, c/\omega_{\rm pe} = \SI{0.096}{\milli\metre}$. The simulations employ a time resolution $\Delta t = 0.0057\, \omega_{\rm pe}^{-1} = \SI{43.7}{\femto\second}$, fulfilling the 3D Courant-Friedrichs-Lewy condition: $c\Delta t <\Delta x/\sqrt{3}$. We employ 8 macro-particles per cell (for each species), and utilize quadratic interpolation with first-order binomial current smoothing. The numerical parameters were carefully chosen after a convergence study with 2D3V PIC simulations.

\clearpage

\backmatter


\providecommand{\noopsort}[1]{}\providecommand{\singleletter}[1]{#1}%

\bmhead{Acknowledgments}
We thank Prof.~C.~Joshi (UCLA), Dr.~F.~Albert (LLNL) and Dr.~C.~Densham (STFC Rutherford Appleton Laboratory) for useful discussions, as well as T.~Levens~(CERN) and Dr.~T.~Ma~(LLNL) for supporting this experiment. This project has received funding from the European Union’s Horizon Europe Research and Innovation programme under Grant Agreement No 101057511 (EURO-LABS).
The work of G.G. was partially supported by UKRI under grants no. ST/W000903/1 and EP/Y035038/1, while A.F.A.B. was also supported by UKRI (grant number MR/W006723/1). The work of P.J.B. and L.O.S. was supported by FCT (Portugal)-Foundation for Science and Technology under the Project X-maser (No. 2022.02230.PTDC) and Grant No. UI/BD/151559/2021. The work of D.H.F. and D.H. was supported by the U.S. Department of Energy under Award Number DE-NA0004144. 
The work of A.A.S. was supported in part by the Simons Foundation via a Simons Investigator Award. 
We also acknowledge funding from AWE plc., and the Central Laser Facility (STFC). 
FLUKA simulations were performed using the STFC Scientific Computing Department's SCARF cluster. OSIRIS simulations were performed at LUMI-C (Finland) within EuroHPC-JU Project
No. EHPC-REG-2021R0038. UK Ministry of Defence © Crown Owned Copyright 2025/AWE.

\bmhead{Author contributions}
This project was conceived by G.G., R.B. and F.M. The experiment was designed by C.D.A., G.G., P.S., N.C. and R.B., and carried out by C.D.A., P.S., N.C., G.G., T.H., R.S., J.W.D.H., P.J.B., S.B., F.D.C., A.G., D.H., S.I., V.S., T.V. and B.T.H. The data analysis was carried out by C.D.A. The manuscript was written by C.D.A., with input from G.G., R.B., N.C., F.M., L.O.S., B.R., P.J.B and S.S. Numerical simulations were performed by C.D.A., P.S, P.J.B. and L.O.S. Further experimental and theoretical support was provided by A.D., I.E., D.H.F., A.F.A.B., A.A.S., J.T.G., B.T.H., S.S., H.C., T.D. and R.M.G.M.T. 

\bmhead{Competing interests}
The authors declare no competing interests.

\bmhead{Supplementary information}
Supplementary Information file provided.




\clearpage

\renewcommand{\figurename}{Supplementary Fig.}
\renewcommand{\tablename}{Supplementary Table}
\renewcommand*\contentsname{\large Contents}
\setcounter{figure}{0}

\section*{\centering{ Suppression of pair beam instabilities in a laboratory analogue of blazar pair cascades: Supplementary Information}}

\vspace{2mm}
\tableofcontents
\vspace{10mm}

\section{FLUKA Monte-Carlo simulations}

\subsection{Modelling the secondary beam generation}\label{sec:supp_MC}

Monte-Carlo simulations of the secondary beam generation were performed using FLUKA to obtain the density and momentum distributions of the most abundantly produced secondary beam components (electrons, positrons, protons and pions). The simplified geometry of the experiment is shown in Supplementary Figure~\ref{fig:FLUKA_model}. The low-energy cutoff for particle transport in the simulation was \SI{10}{\kilo\electronvolt} for $\rm{e}^-/\rm{e}^+/\gamma$ and \SI{100}{\kilo\electronvolt} for hadrons. $10^5$ primary protons were transported through the geometry, leading to agreement between the calculated and measured beam properties on the order of $1\%$ in the regions of interest, as discussed in Ref.~\cite{arrowsmith2023laboratory}.

\begin{figure}[h]
\centering
\includegraphics[width=1\textwidth]{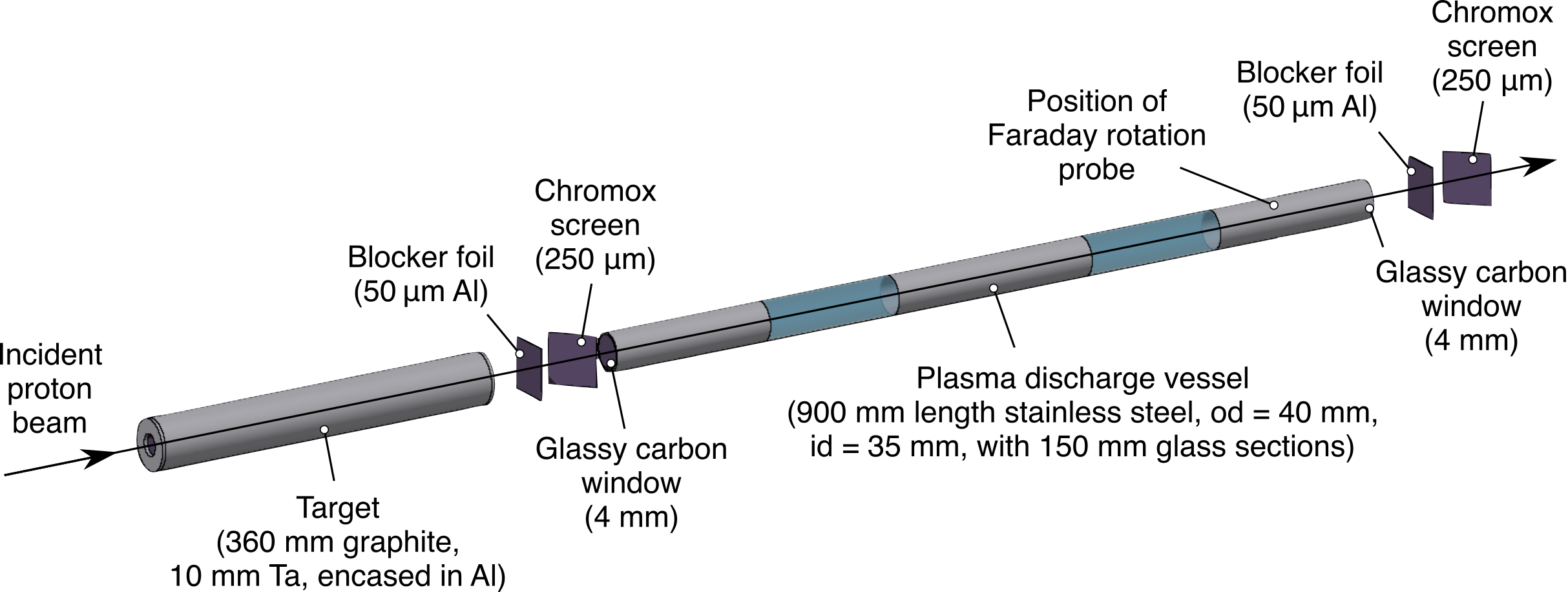}
\caption{{\bf Simplified geometry of the experiment simulated using FLUKA.} The plasma discharge vessel is modelled as a stainless steel tube with two glass sections where the inductive coils are located. The surrounding experimental area including the downstream beam dump as well as the concrete walls and shielding are also included in FLUKA simulations (not shown). The exact material compositions are used in all cases. 
}
\label{fig:FLUKA_model}
\end{figure}

\clearpage
\subsection{Obtaining a value for $B_0$}

To obtain an upper bound on the instability's growth rate, the maximum inferred magnetic field measurement from the Faraday rotation diagnostic is compared with the expected measurement when no plasma is present. Given the density distribution of secondary particles at the probe position in a FLUKA simulation, the spatial distribution of the net current is used to calculate the magnetic field via Amp\`ere's circuital law:
\begin{equation}
    \oint_C\mathbf{B}\cdot\mathrm{d}\mathbf{l} = \mu_0\iint_S \mathbf{J}\cdot\mathrm{d}\mathbf{S} = \mu_0 I_{\mathrm{enc}},
\end{equation}
where $I_{\mathrm{enc}}$ is the current passing through an enclosed surface space, $S$, bounded by contour, $C$. The magnitude of the azimuthal magnetic field as a function of radius is shown in Supplementary Figure~\ref{fig:B_phi}. The estimated magnetic field measurement in the absence of beam-plasma interaction is calculated from $B_{\phi}(r)$.

\begin{figure}[h]
\centering
\includegraphics[width=1\textwidth]{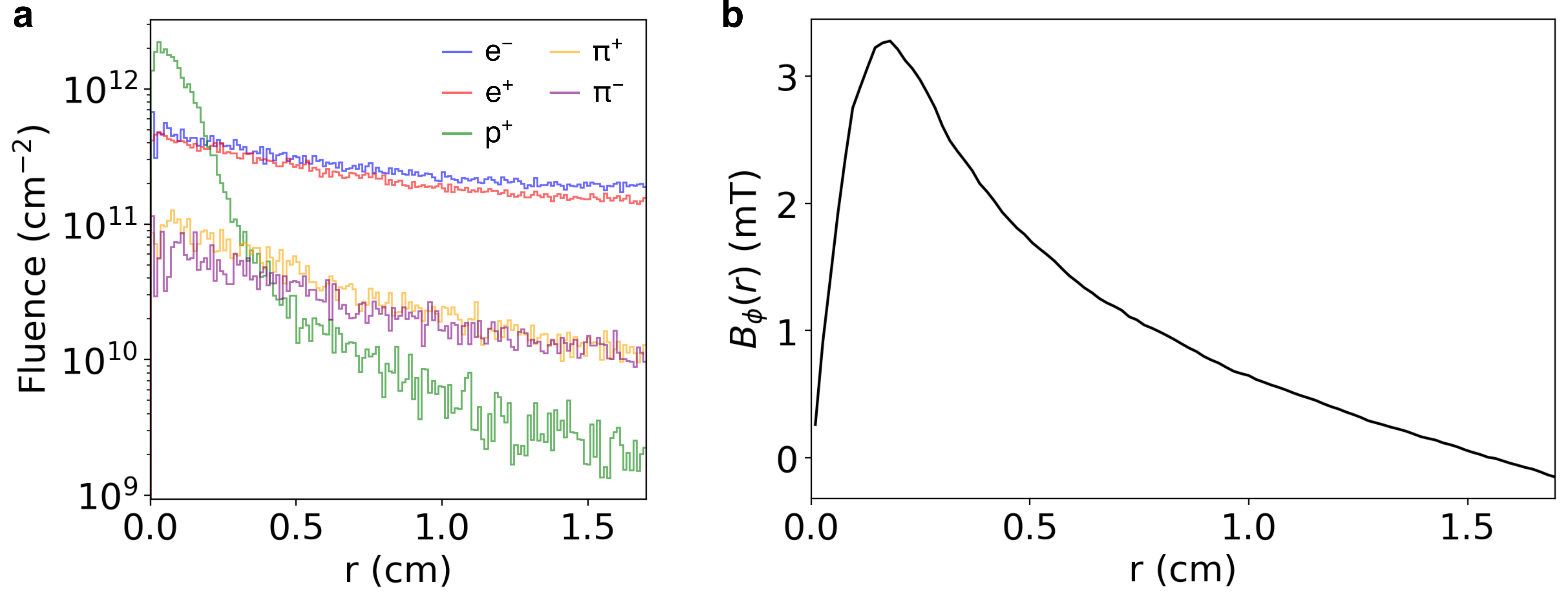}
\caption{{\bf Obtaining the azimuthal magnetic field at the Faraday probe position in the absence of plasma using FLUKA simulations.} (a) The particle fluence is plotted as a function of radius at the position corresponding to the Faraday probe inside the plasma vessel (obtained from a FLUKA simulation) for electrons (blue), positrons (red), protons (green), positive pions (orange) and negative pions (purple). (b) Amp\`ere's circuital law is used to calculate the corresponding azimuthally-oriented magnetic field as a function of radius.
}
\label{fig:B_phi}
\end{figure}

\clearpage
\subsection{Fitting analytical forms to the momentum phase space of secondary beam components}\label{sec:supp_PIC}

\begin{figure}[h]
\centering
\includegraphics[width=1\textwidth]{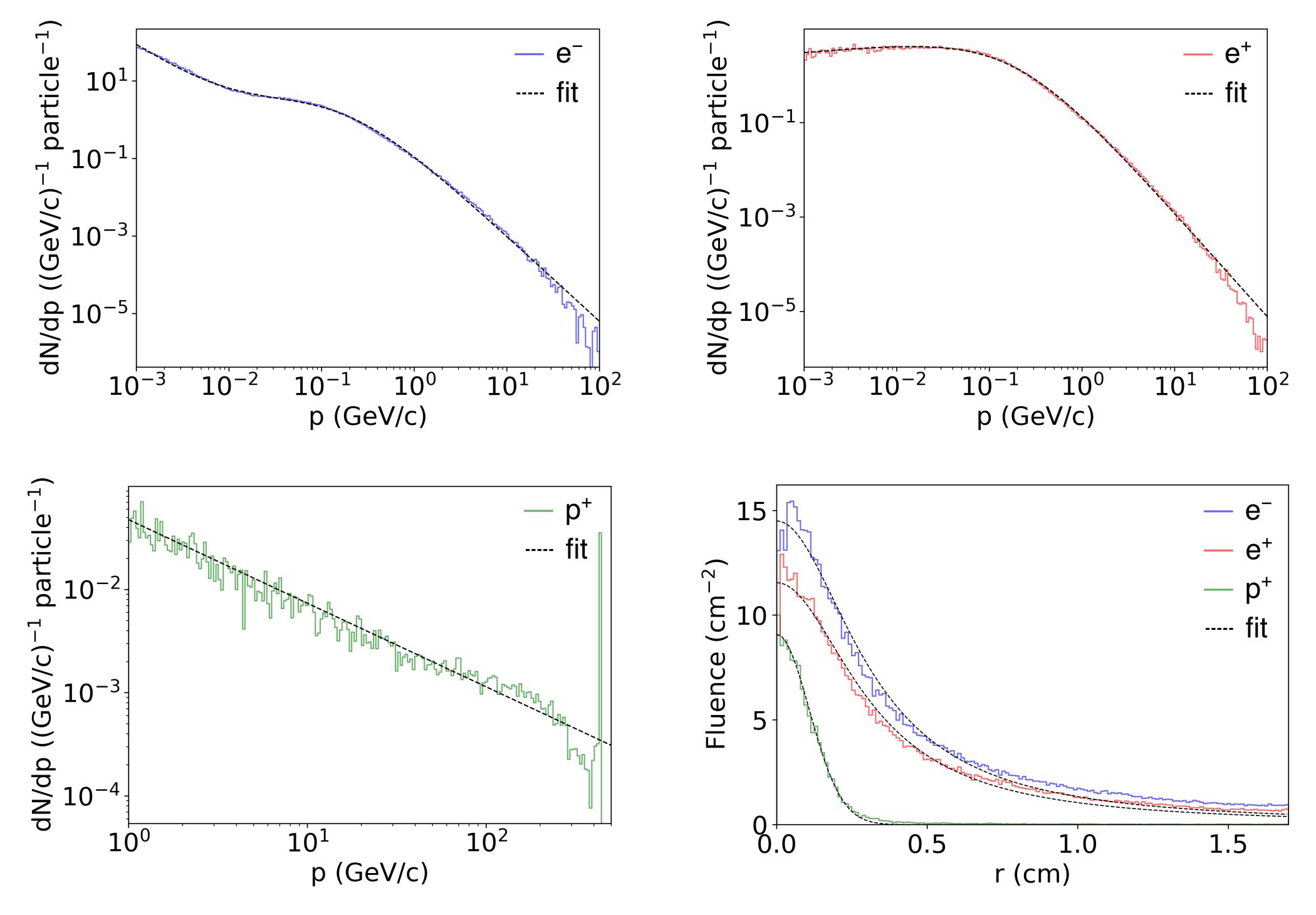}
\caption[Fitted momentum and density distributions of beam species]{{\bf Fitted momentum and density distributions of beam species.} The momentum and density distribution of electrons (blue), positrons (red) and protons (green), per incident primary proton, at the entrance of the plasma (after passing through the glassy carbon window), obtained from a FLUKA simulation and fitted to analytical forms (black-dashed).
}
\label{fig:PIC_fitting_1}
\end{figure}

\begin{figure}[h]
\centering
\includegraphics[width=0.8\textwidth]{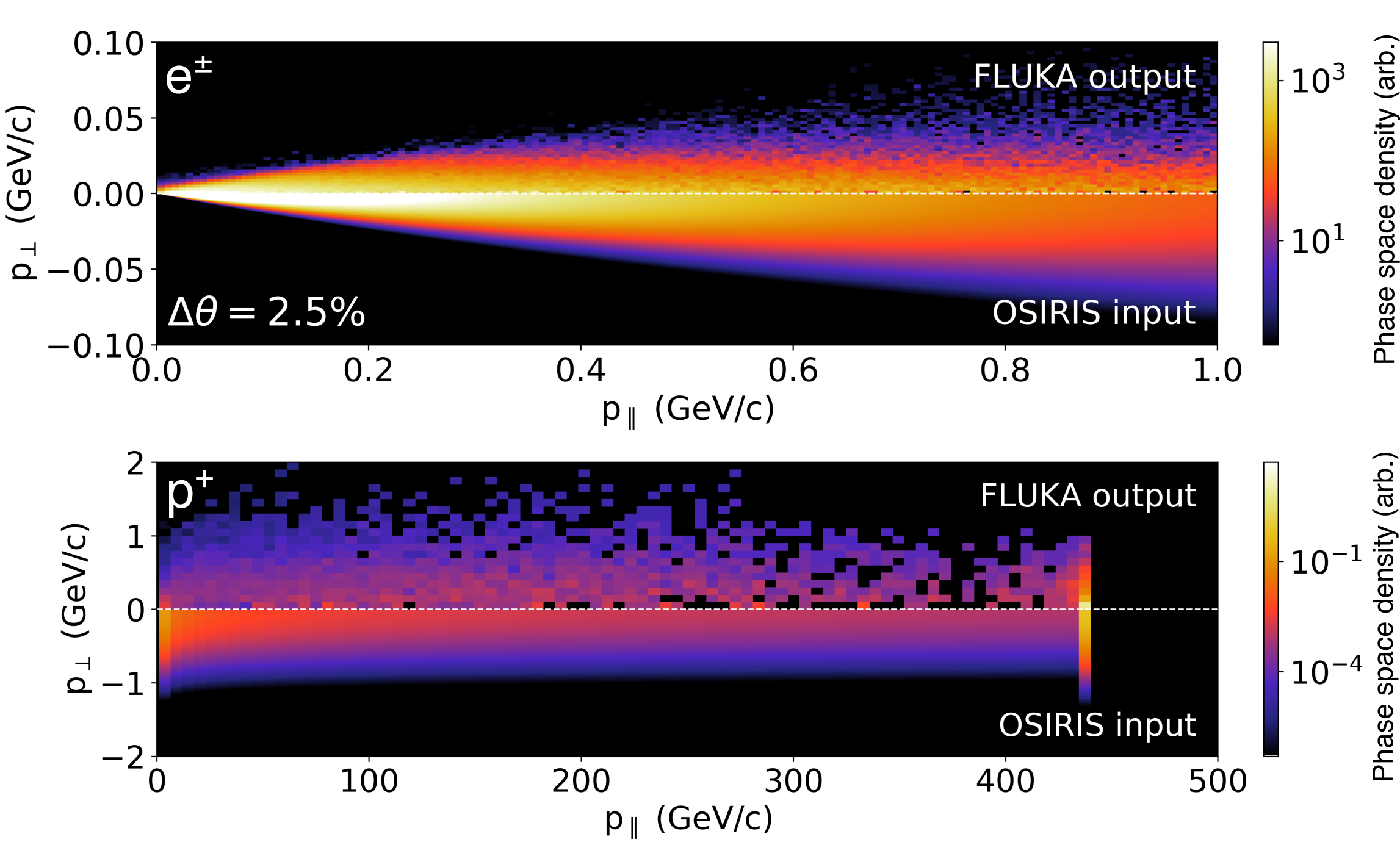}
\caption[Fitted momentum-space distributions of beam species]{{\bf Fitted momentum-space distributions of beam species.} The fits to the full momentum-space distribution are shown for e$^{\pm}$ (upper) and p$^{+}$ (lower). In each frame, the top half shows the momentum space obtained from a FLUKA simulation, and the lower half shows the analytical fitted function. For the e$^{\pm}$ pairs, a Gaussian-distributed transverse momentum with $\sigma_{p_{\perp}} = p_{\parallel}\cdot\Delta\theta$ and $\Delta\theta = 0.025$ provides an appropriate fit. For the p$^{+}$, $\sigma_{p_{\perp}}=\SI{0.25}{\giga\electronvolt}/c$.
}
\label{fig:PIC_fitting_2}
\end{figure}

From FLUKA simulations, the proton spectrum is split into the population that undergoes no significant inelastic scattering ($N_{\rm{p}^+}=1.26\times10^{11}$, represented by a delta function in momentum space at $p=\SI{440}{\giga\electronvolt}/c$) and the population at lower momenta ($p<\SI{440}{\giga\electronvolt}/c$) generated in the hadronic cascade, which has a power-law momentum spectrum:
\begin{equation}
\frac{dN}{dp} = AN_{\rm{p}^+}p^{k_1},
\end{equation}
where $N_{\rm{p}^+}=1.50\times10^{11}$, $k_1=-0.81$, and the prefactor $A$ 
is fixed by the normalization condition $\int_{x_{\rm min}}^{x_{\rm max}} \left(\frac{dN}{dp}\right)\,dp = 1$. The proton transverse momentum spread is Gaussian with a standard deviation $\sigma_{p_{\perp}}=\SI{0.25}{\giga\electronvolt}/c$. The density profile is Gaussian distribution in all dimensions:
\begin{equation}
\rho(r,z) = \frac{1}{(2\pi)^{3/2}\sigma_r^2\sigma_z} \exp\left[-\left(\frac{r^2}{2 \sigma_r^2} + \frac{z^2}{2 \sigma_z^2}\right)\right], 
\end{equation}
with $\sigma_r=\SI{0.11}{\centi\metre}$ and $\sigma_z=\SI{7.5}{\centi\metre}$. Electrons and positrons are modelled in the spectral range $\SI{1}{\mega\electronvolt}-\SI{10}{\giga\electronvolt}$ with a multi-power-law distribution. The electron momentum spectrum is modelled by a triple-index power-law distribution:
\begin{equation}
\frac{dN}{dp} = AN_{\rm{e}^-}p^{k_1}(p+p_{1,2})^{k_2-k_1}(p+p_{2,3})^{k_3-k_2},
\end{equation}
where $N_{\rm{p}^-}=5.48\times10^{12}$, $k_1=-1.6$, $p_{1,2}=\SI{0.012}{\giga\electronvolt}$, $k_2=0.8$, $p_{2,3}=\SI{0.25}{\giga\electronvolt}$, $k_3 = -2.2$, and the prefactor $A$ 
is fixed by the same normalization condition: 
\begin{equation}
\begin{gathered}
    A~\bigg[(k_1+1)^{-1} p^{k_1+1}(p+p_{1,2})^{k_2-k_1}\bigg(\frac{p+p_{1,2}}{p_{1,2}}\bigg)^{k_1-k_2}(p+p_{2,3})^{k_1-k_3}  \bigg(\frac{p+p_{2,3}}{p_{2,3}}\bigg)^{k_3-k_1} \\ F_1\bigg(k_1+1,k_1-k_2,k_3-k_1;k_1+2;-\frac{p}{p_{1,2}},-\frac{p}{p_{2,3}}\bigg)\bigg]_{p_{\rm min}}^{p_{\rm max}}=1,
\end{gathered}
\end{equation}
where $F_1$ is the Appell hypergeometric function. The positron momentum spectrum is modelled by a double-index power-law distribution:
\begin{equation}
\frac{dN}{dp} = AN_{\rm{e}^+}p^{k_1}(p+p_{1,2})^{k_2-k_1},
\end{equation}
where $N_{\rm{e}^+}=4.27\times10^{12}$, $k_1=0.17$, $p_{1,2}=\SI{0.19}{\giga\electronvolt}$, $k_2=-2.2$, and $A$ is determined by
\begin{equation}
    A~\Bigg[\Bigg(\frac{p^{k_1+1}p_{1,2}^{k_2-k_1}}{k_1+1}\Bigg)~_2F_1\bigg(k_1+1,k_1-k_2;k_1+2;-\frac{p}{p_{1,2}}\Bigg)\bigg]_{p_{\rm min}}^{p_{\rm max}} = 1,
\end{equation}
where $_2F_1$ is the hypergeometric function. The density profiles follow a Gaussian distribution longitudinally and a Cauchy distribution radially:
\begin{equation}
\rho(r,z) = \frac{1}{(2\pi)^{3/2}\sigma_z} \frac{\frac{1}{2}\Sigma_r}{(r^2+(\frac{1}{2}\Sigma_r)^2)^{3/2}} \exp\left(-z^2/2 \sigma_z^2\right), 
\end{equation}
with $\sigma_z=\SI{7.5}{\centi\metre}$ and a radial half-width-half-maximum $\Sigma_r=\SI{0.32}{\centi\metre}$. In Eq.~\ref{eq:oblique}, the transverse momentum spread of pairs is defined $\Delta\theta \equiv \Delta p_{\perp}/p_{\parallel} = \big[\frac{1}{n_{\pm}}\int p_{\perp}^2f_0\,d\mathbf{p}\big]^{1/2}/\gamma_{\pm}m_{\mathrm{e}}c$, where $f_0$ is the pair distribution function normalized by $n_{\pm}=\int f_0\,d\mathbf{p}$. To obtain a value for $\Delta \theta$, the transverse momentum spread of pairs is fit to a Gaussian with a standard deviation that is a function of the longitudinal momentum: 
\begin{equation}
\frac{d^2N}{dp_{\perp}^2} = \frac{1}{2\pi\sigma_{p_{\perp}^2}}\exp\left(-p_{\perp}^2/2 \sigma_{p_{\perp}}^2\right),\quad \sigma_{p_{\perp}} = p_{\parallel}\cdot\Delta\theta,
\end{equation} 
where $\Delta\theta = 0.025$ provides an appropriate fit (shown in Figures~\ref{fig:PIC_fitting_1} and \ref{fig:PIC_fitting_2}).

\clearpage

\section{Particle-in-cell simulations}

\begin{figure}[b!]
\centering
\includegraphics[width=0.95\textwidth]{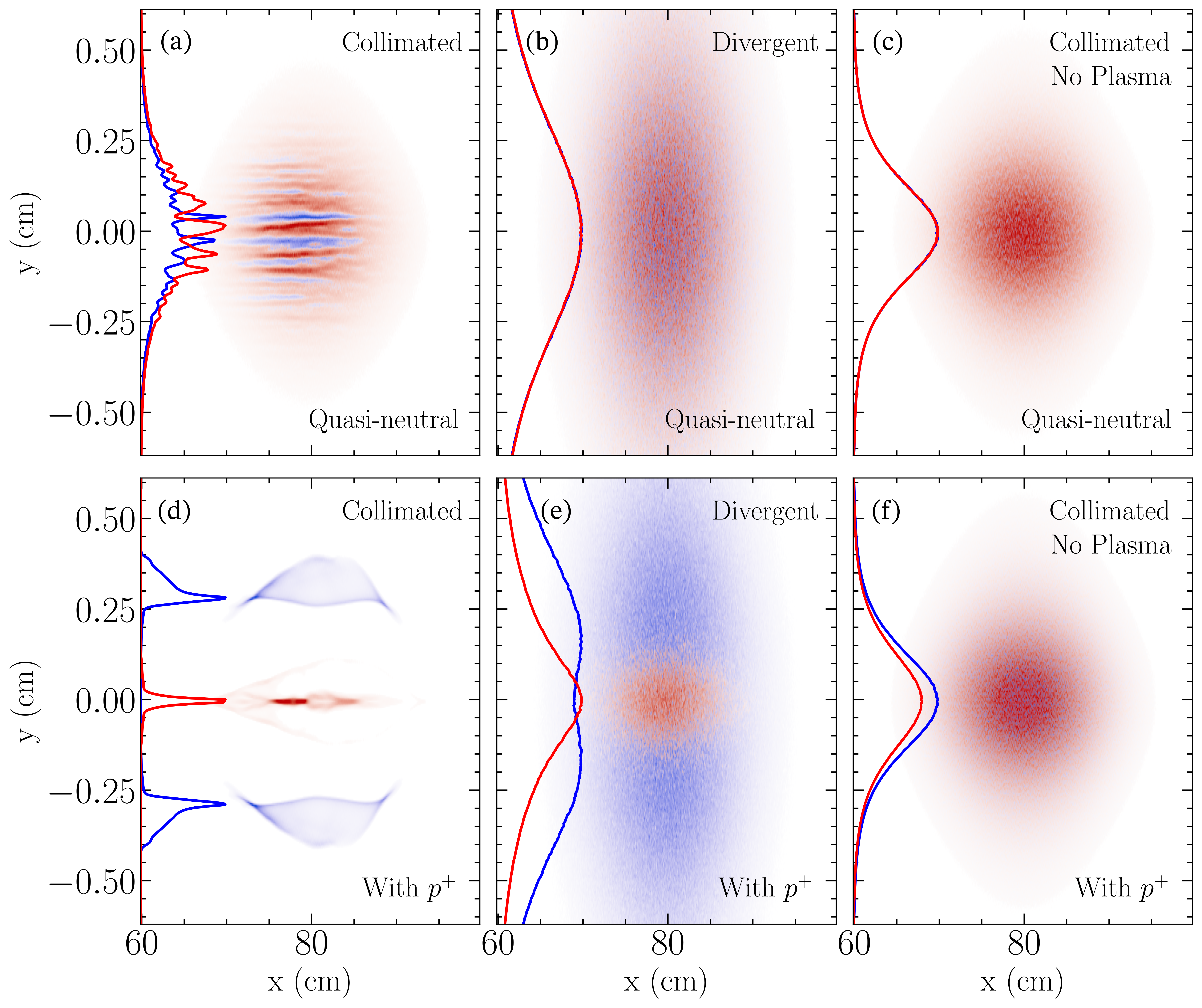}
\caption[Additional 2D particle-in-cell simulations.]{{\bf Additional 2D particle-in-cell simulations.} \textbf{(a-c)} Without protons and quasi-neutral pairs: (a) collimated pairs, (b) diverging pairs (experimental beam divergence), (c) collimated pairs, without plasma. \textbf{(d-f)} With protons and realistic electron-positron ratio: (d) collimated pairs, (e) diverging pairs, (f) collimated pairs, without plasma. All snapshots are taken when the beam is at the location of the Faraday probe.
}
\label{fig:PIC_supp_1}
\end{figure}

As shown in Figure 3 of the main paper, the growth rate that is obtained in PIC simulations of a collimated pair beam with co-propagating residual protons as in the experiment, agrees with the predictions from quasi-linear theory (Eq. 1). Since Eq. 1 is derived assuming a quasi-neutral beam (with no protons), we do not expect that the presence of the residual proton beam has any other dynamical effects, except for providing a seed for the instabilities' growth. To confirm this, we have performed six additional 2D simulations of the experimental setup, shown in Supplementary Figure~\ref{fig:PIC_supp_1}. The top row corresponds to simulations that are fully quasi-neutral, no proton beam and the same number of electrons and positrons. The bottom row shows simulations including the residual proton beam and the experimental ratio of electrons to positrons. The different columns correspond to simulations where the beam is collimated (first column), the beam has the experimental beam divergence (second column), and without a background plasma (third column). The collimated, quasi-neutral (QN) case shows clear filaments like the usual current filamentation instability. For the divergent, quasi-neutral case we do not observe any signs of the instability, and similarly without a plasma, no instability develops. Once the proton beam is included, we obtain results similar to the 3D simulations in the main text. The presence of the proton beam seeds the instability (this can be seen in the temporal evolution of the spectrum of spatial fluctuations in the fireball beam or in the current, for instance). The seeding determines from where the instability is growing, but the growth behaviors (notably the suppression and the respective growth rates) are comparable, as seen in Supplementary Figure~\ref{fig:PIC_supp_2}.

Finally, there is the question of whether this is an instability dominated by the magnetic field or just the electrostatic fields. For this we can look at the total energy that is converted into the $E$ and $B$-fields (see Supplementary Figure~\ref{fig:PIC_supp_3}). Here, the energy in the simulation domain is normalized to the initial beam energy (hence the small numbers). In the top plot, the beam is quasi-neutral. The divergent case initially grows due to noise, but not much increase is observed after that, there are some spikes in the $E$-field energy due to the absence of plasma not being able to screen any charge separation. In contrast, the collimated case has $E>B$, until the instability triggers, at which point $B>E$ and clear growth is observed, meaning that the beam is magnetically dominated, i.e., there is a frame, the so-called Weibel frame, where $E=0$ and there is only a $B$-field~\cite{lemoine2019physics}. Comparing with the simulations including protons, the divergent case starts at a lower normalized energy because the hotter beam has more kinetic energy. But, similar trends are observed when compared with the collimated case. The $E$-field is rapidly screened as the beam enters the plasma, and as it propagates there is a dip in the $B$-field due to the background plasma establishing a return current. This is followed by an increase of the $B$-field due to the instability, and as the beam propagates in the cell and enters the middle region (with lower density) the $E$-field increases with the $B$-field as the plasma does not efficiently screen the currents and charges of the beam. 
Then similar dynamics occur and at the exit of the plasma both beams end with $B>E$ indicating magnetization of the beam.

\begin{figure}[h!]
\centering
\includegraphics[width=0.85\textwidth]{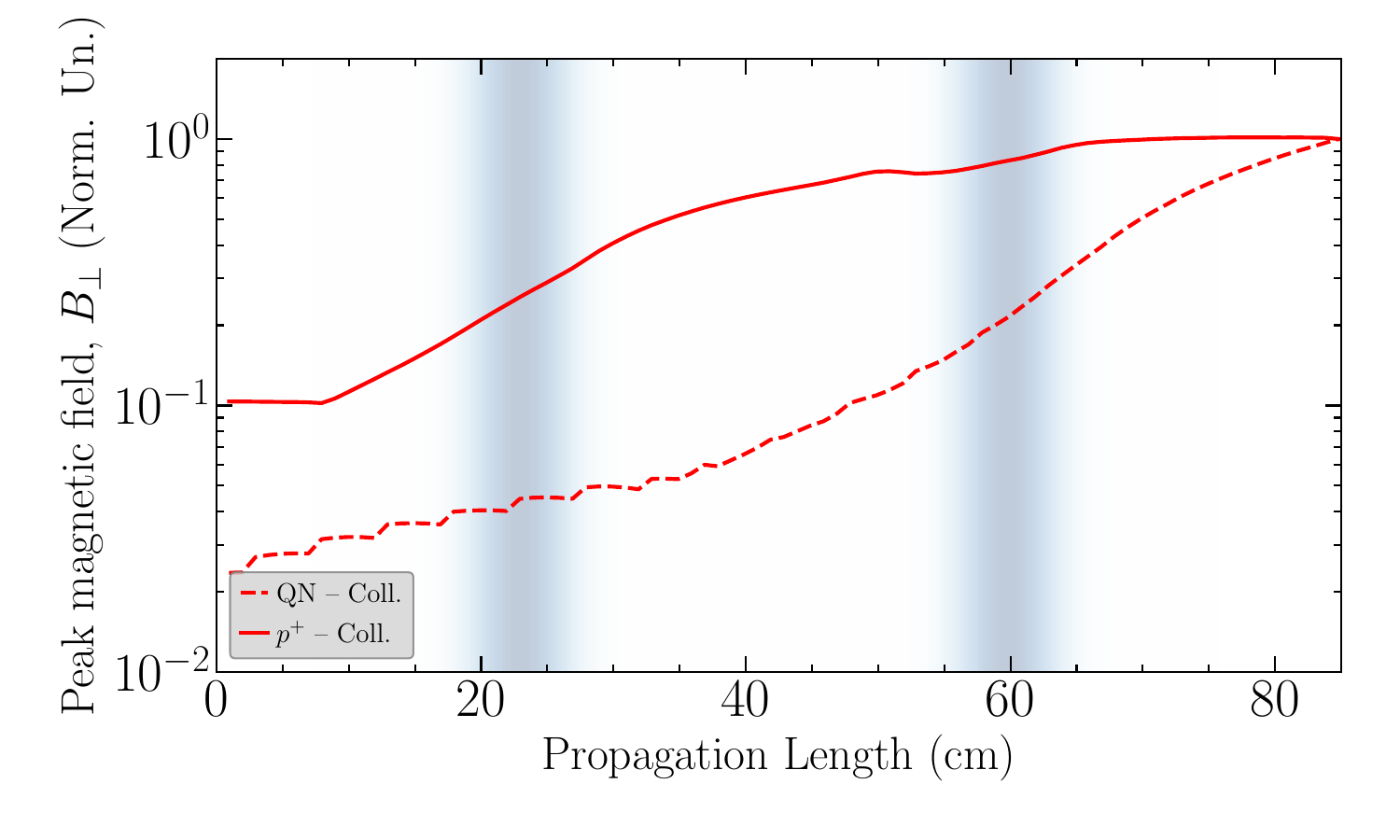}
\caption[]{{\bf Growth of the peak magnetic field in additional particle-in-cell simulations.} Peak magnetic field is plotted as function of propagation distance of the beam through the plasma cell assuming collimated pairs. `QN' refers to simulations without protons and quasi-neutral pairs (dashed red), and `$p^+$' refers to simulations where the proton beam is included and the experimental ratio of electrons to positrons is assumed (solid red). The `propagation length' is defined as the position of the beam relative to when the beam's center is at the entrance of the plasma cell. Shaded regions correspond to the regions of higher plasma density, as in main text Figure~\ref{fig:3DPIC}c.}
\label{fig:PIC_supp_2}
\end{figure}

\begin{figure}[h!]
\centering
\includegraphics[width=0.85\textwidth]{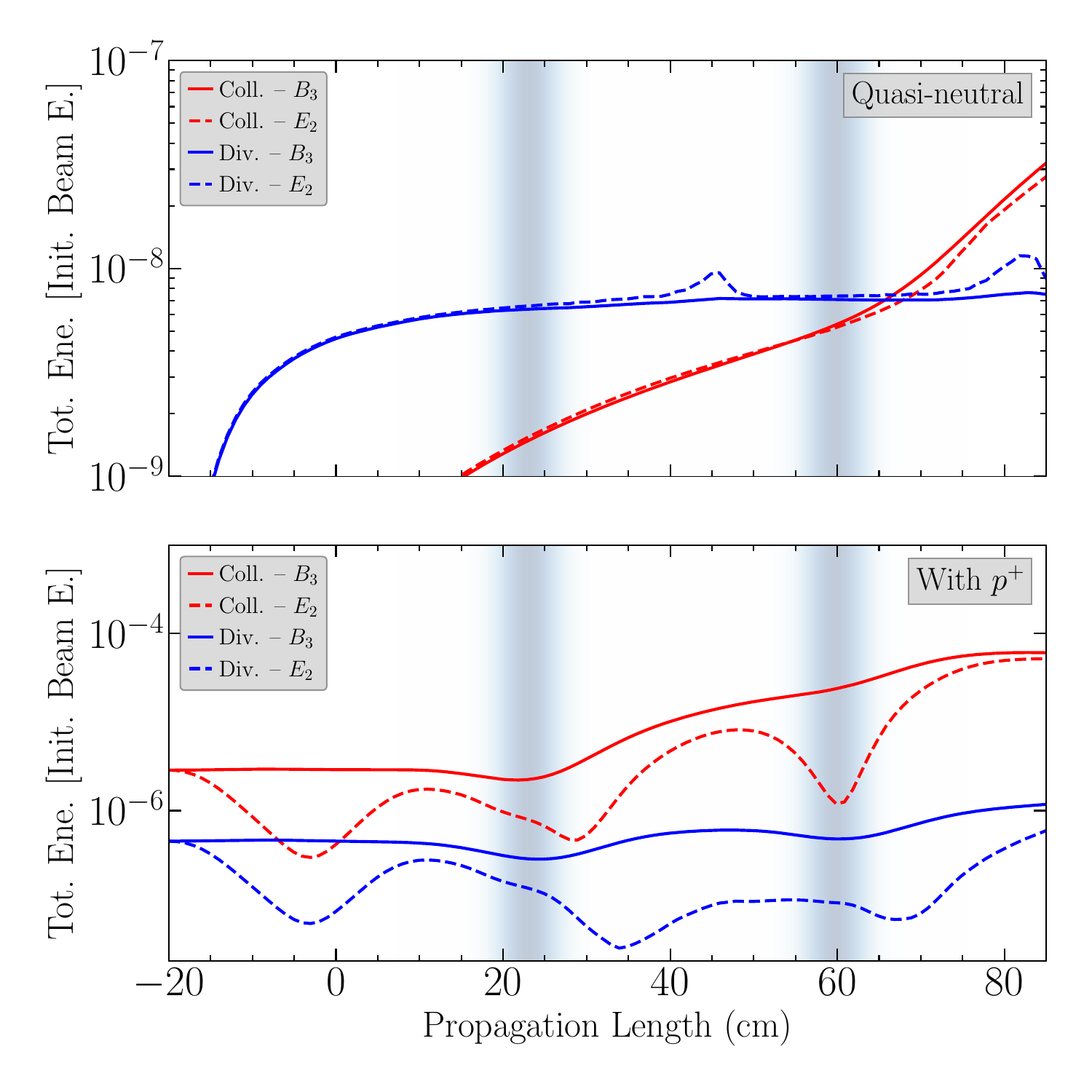}
\caption[]{{\bf Growth of electric and magnetic field energy in additional particle-in-cell simulations.} Transverse magnetic ($B_3$, solid) and electric ($E_2$, dashed) fields are plotted assuming collimated (red) and diverging (blue) pairs assuming quasi-neutral pairs without protons (top) and including protons and the experimental ratio of electrons to positrons (bottom). Energies are normalized to the total beam energy, including the bulk kinetic energy. `Propagation length' is the same as defined in the caption of Supplementary Figure~\ref{fig:PIC_supp_2}.
}
\label{fig:PIC_supp_3}
\end{figure}

\clearpage

\section{Experimental measurements}

\subsection{Langmuir probe measurements of the plasma electron temperature and density profiles}\label{sec:Langmuir}

The plasma is characterized extensively for a range of argon gas fill pressures ($p_{\mathrm{g}}$) and absorbed powers ($P_{\mathrm{abs}}$) prior to the experiment using a Langmuir probe. A commercially available Langmuir probe unit was used (manufactured by Impedans, Ireland) with a single probe tip configuration in combination with an RF compensation electrode. The probe tip is made of tungsten wire (length $\SI{4}{\milli\metre}$, radius $\SI{0.195}{\milli\metre}$), held in position by an alumina tube (radius $\SI{2}{\milli\metre}$). The probe tip is connected to the data acquisition unit via a series of RF chokes to filter RF interference. The probe is inserted into the plasma at different positions along the axis to obtain the electron density, ion density, electron temperature and energy distribution function. The density and temperature profiles measured from the centre of the coil to the end of the discharge are shown in Supplementary Figure~\ref{fig:Langmuir_probe_measurements}, for absorbed powers: $P_{\rm abs}=\SI{175}{\watt},\SI{200}{\watt},\SI{225}{\watt}$, and fill pressures: $p_{\rm g}=\SI{1}{\pascal}$ and \SI{4}{\pascal}. The highest plasma densities are measured when the probe tip is positioned under the centre of the inductive coil. At any given pressure, the peak plasma density scales approximately linearly with the absorbed power. In particle-in-cell simulations the longitudinal plasma density profile is modelled by:
\begin{equation}
n_{\mathrm{p}}(z)= \frac{n_0\left[ e^{-(z-z_1)^2/(2 \sigma_c^2)}+ e^{-(z-z_2)^2/(2 \sigma_c^2)} + n_{\mathrm{min}}\right]}{\left(1+e^{-z/20}\right) \left(1+e^{-(-z+l)/20}\right)},
\end{equation}
where $\sigma_{\rm c}=\SI{3.75}{\centi\metre}$, $l=\SI{87}{\centi\metre}$, $n_\mathrm{min}=0.02$, and $z_1=\SI{24.88}{\centi \metre}$ and $z_2= \SI{61.88}{\centi\metre}$ correspond to the centres of the inductive coils. These parameters correspond to argon fill pressure $p_{\mathrm{g}}=\SI{4}{\pascal}$ and absorbed power $P_{\mathrm{abs}}=240\pm\SI{10}{\watt}$ (shown in Figure~\ref{fig:Experimental_Setup}c of the main text).

\begin{figure}[h]
\centering
\includegraphics[width=1\textwidth]{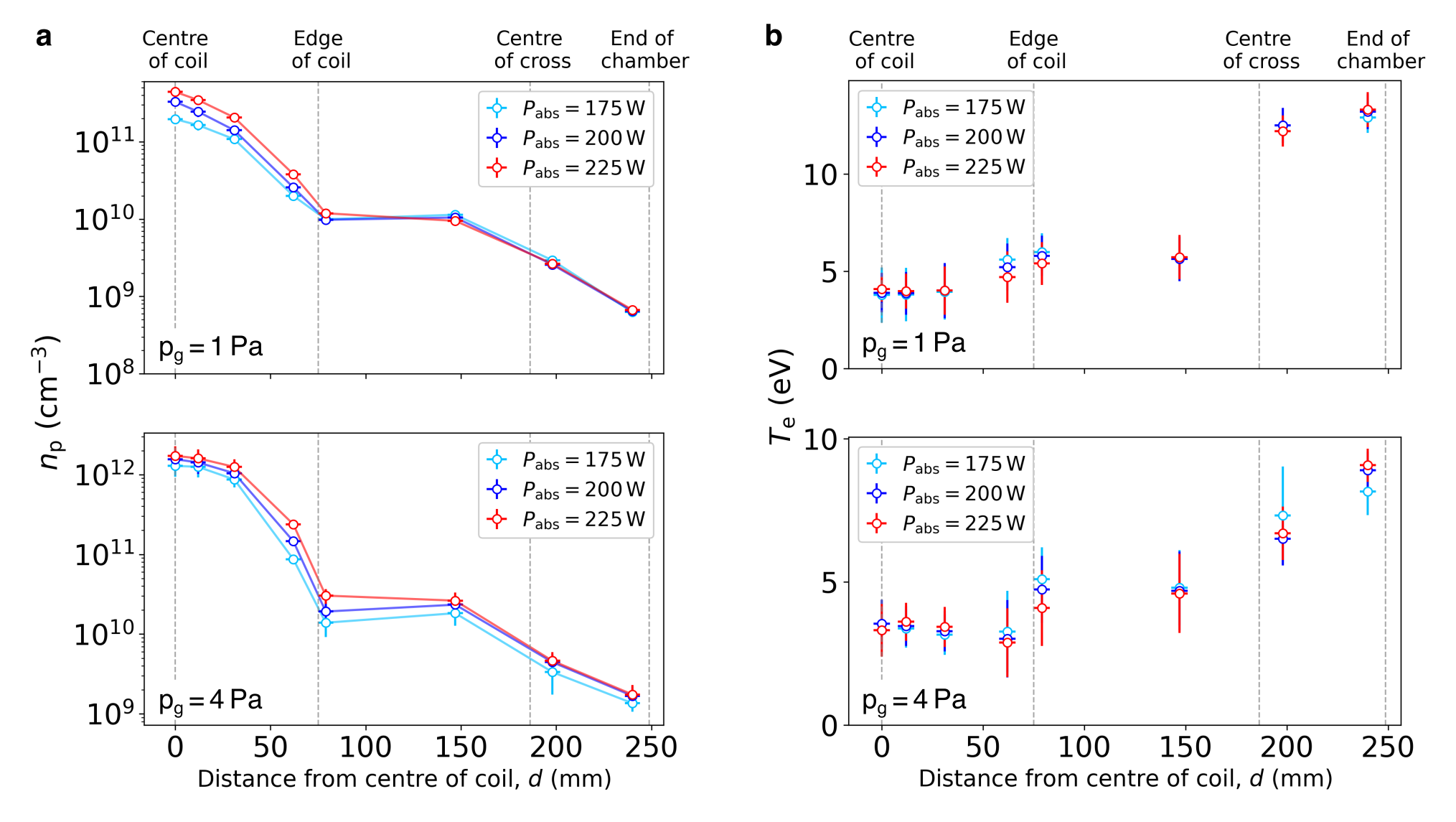}
\caption{{\bf Langmuir probe measurements of the plasma density and temperature.} (a) The plasma density, $n_{\rm p} = (n_{\rm e}+ n_{\rm i})/2$ and (b) electron temperature, $T_{\rm e}$, measured at different axial positions along the discharge relative to the centre of one of the inductive coils. Three power settings and pressure settings are measured: $P_{\rm abs}=\SI{175}{\watt},\SI{200}{\watt},\SI{225}{\watt}$, and fill pressures: $p_{\rm g}=\SI{1}{\pascal}$ and \SI{4}{\pascal}. The vertical dashed lines correspond to: the centre of the coil ($z = \SI{0}{\milli\metre}$), the edge of the coil ($z = \SI{75}{\milli\metre}$), the centre of the six-way cross ($z = \SI{186}{\milli\metre}$), and the end of the chamber ($z = \SI{245}{\milli\metre}$).
}
\label{fig:Langmuir_probe_measurements}
\end{figure}

\clearpage

\subsection{Optical emission spectroscopy of the argon plasma}\label{sec:OES}

A Langmuir probe is too invasive to perform in-situ measurements during the experiment, so optical spectrographs of the plasma emission are collected and the spectra are compared with those obtained prior to the experiment simultaneously with Langmuir probe measurements. Spectroscopic measurements were performed using a Princeton Instruments Acton SpectraPro Czerny-Turner spectrometer coupled to a Hamamatsu Orca flash camera (with a \SI{2}{\second} exposure time). Light was collected from a $\SI{2}{\milli\metre}$ area at the centre of the inductive coil using a collimating lens, which coupled the light into an optical fibre and transported it to the slit of the spectrometer. The optical fibre is multi-mode with core diameter $\SI{105}{\micro\metre}$ and numerical aperture 0.1. The slit width was $\SI{70}{\micro\metre}$ and the grating had a line density 600\,g/mm. A calibrated white light source was used to account for any spectral dependence on transmission/reflection of the optics or spatial variation in the camera sensitivity. An argon-mercury lamp was used to calibrate the spectrum and obtain the instrument function. The spectra were measured in the spectral region $\lambda = 725-770~\si{\nano\metre}$, where emission lines are prominent and variation in the line ratios is observed for different plasma conditions. Raw data and lineouts for two sets of plasma conditions are shown in Supplementary Figure~\ref{fig:OES_example}. Emission from argon in the neutral state Ar(I) dominates the observed emission due to the low ionization fraction.

\begin{figure}[h]
\centering
\includegraphics[width=0.8\textwidth]{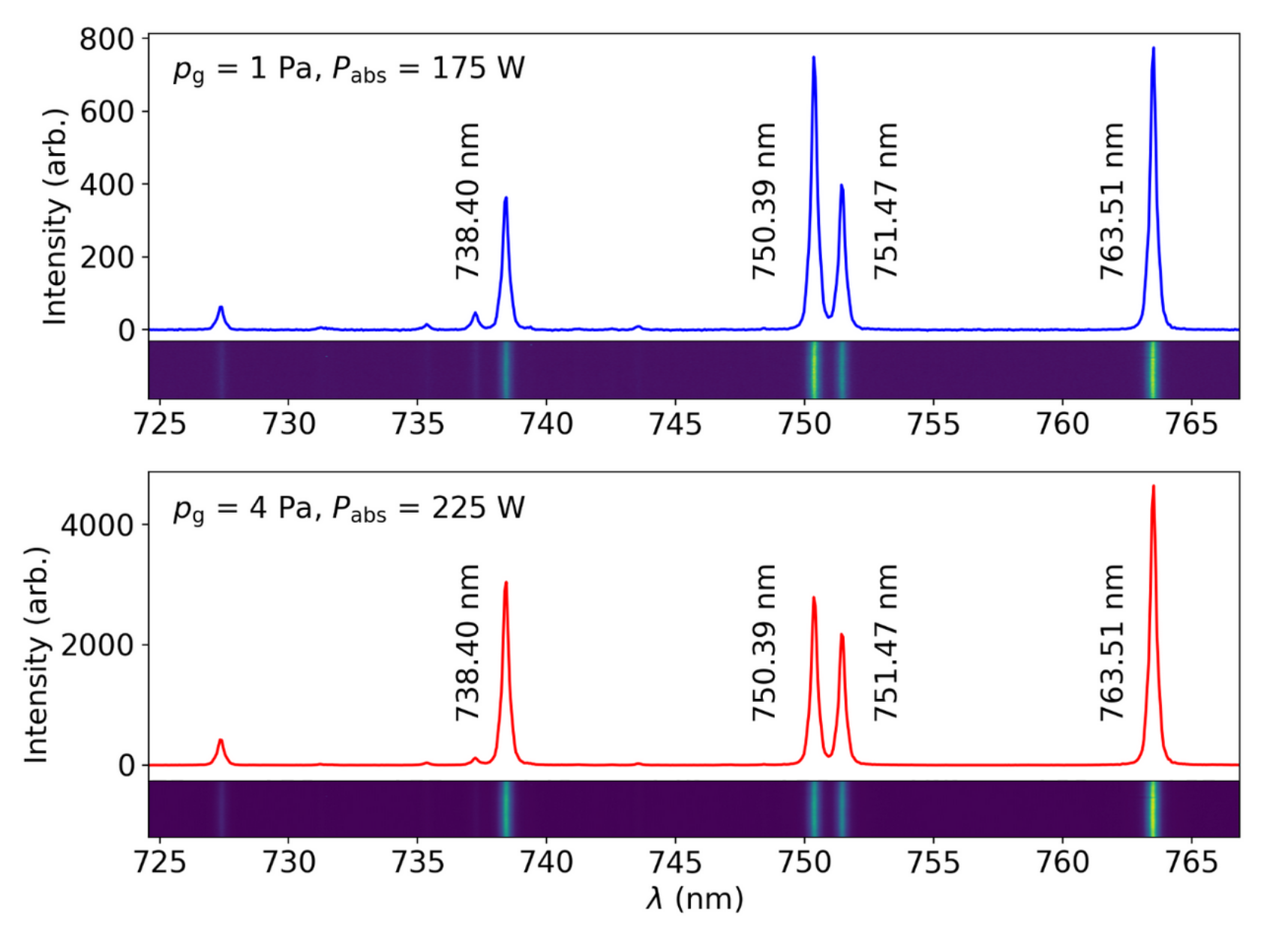}
\caption{{\bf Optical emission spectroscopy of the argon plasma.} The optical emission from the plasma at the centre of the inductive coil is measured in the spectral range $\lambda =725-770$~nm for two sets of plasma conditions. Upper panel: $p_{\mathrm{g}}=\SI{1}{\pascal}$, $P_{\mathrm{abs}}=\SI{175}{\watt}$, corresponding to a plasma electron density $n_{\mathrm{e}}=(2.0\pm0.2)\times\SI{e11}{\per\centi\metre\cubed}$ and temperature $T_{\mathrm{e}}=3.8\pm1.4\,\si{\electronvolt}$ measured before the experiment using a Langmuir probe. Lower panel: $p_{\mathrm{g}}=\SI{1}{\pascal}$, $P_{\mathrm{abs}}=\SI{175}{\watt}$, corresponding to a plasma electron density $n_{\mathrm{e}}=(1.7\pm0.1)\times\SI{e11}{\per\centi\metre\cubed}$ and temperature $T_{\mathrm{e}}=3.3\pm0.9\,\si{\electronvolt}$.
}
\label{fig:OES_example}
\end{figure}

\clearpage

\subsection{Design and characterization of the Faraday rotation probe}\label{sec:Faraday}

The setup of the time-resolved magneto-optic Faraday rotation probe is shown in Supplementary Figure~\ref{fig:Faraday_layout}. The inset in Supplementary Figure~\ref{fig:Faraday_layout} also shows the arrangement of the Verdet crystal inside the plasma. Measurements of the Verdet constant, intrinsic noise and instrument response function are shown in Supplementary Figure~\ref{fig:verdet_performance}. To calibrate the Verdet constant, the change in probe beam intensity ($\Delta V$) is recorded when the TGG crystal is exposed to a permanent magnetic field (neodymium ring magnet K\&J Magnetics RZ0Y0X0, 3" outer diameter, 2" inner diameter, 1" thickness). The field strength of the magnet is measured using a Hall probe. 
The intensities of the orthogonal polarization components change sinusoidally with the applied magnetic field strength, giving a Verdet constant $\mathcal{V}=217\pm\SI{15}{\radian\per\tesla\per\metre}$. The intrinsic electronic noise (of the photodiode and oscilloscope combination) is Gaussian-distributed with a standard deviation $\sigma_{\rm n1} = \SI{0.354}{\milli\volt}$. For $N$ multiple shots, the standard deviation of the mean signal is taken as $\sigma_{\mathrm{n}}=\sigma_{\mathrm{n}1}/\sqrt{N}$. The instrument response function is measured using mV-scale impulse signals produced by exposing the photodiode to an attenuated ultra-short femtosecond laser. The measured rise time is $t_{10-90\%}=\SI{0.44}{\nano\second}$, and the instrument response function convolved with the temporal profile of the primary proton beam (Gaussian with $\sigma_t=\SI{0.25}{\nano\second}$, rise time $t_{10-90\%}=\SI{0.42}{\nano\second}$) is shown in Supplementary Figure~\ref{fig:verdet_performance}, demonstrating that changes in intensity on timescales of the beam duration can be detected. The magnetic field that corresponds to an observed change in probe beam intensity is shown in Supplementary Figure~\ref{fig:Faraday-diagnostic_performance}. The 1-$\sigma$ intrinsic noise results in an instrument sensitivity of $ B_{\rm sens}=\SI{5}{\milli\tesla}$.

\begin{figure}[htp]
\centering
\includegraphics[width=0.95\textwidth]{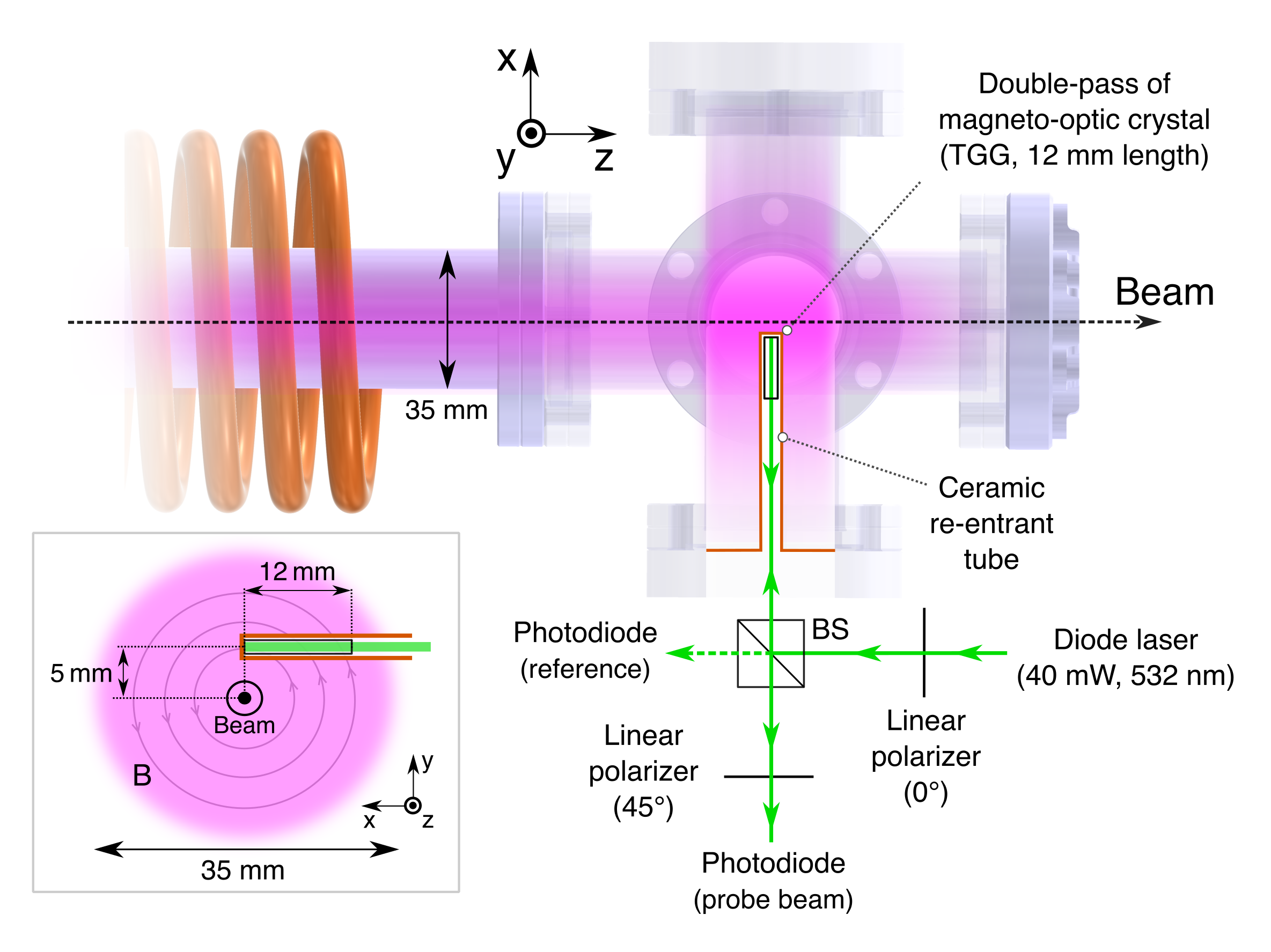}
\caption[Design of the time-resolved magneto-optic Faraday probe]{{\bf Design of the time-resolved magneto-optic Faraday probe.} Green diode emission ($\lambda=\SI{532}{\nano\metre}$) is transported via an optical fibre to the position of the Faraday probe at the end of the plasma discharge. The laser is linearly polarized before it is passed through a 50:50 beamsplitter (BS). The transmitted beam (coupled into a fibre and measured using a photodiode) acts as a reference to changes in laser intensity. The reflected component is directed along a ceramic re-entrant tube to a TGG magneto-optic crystal suspended in the plasma. The rear surface of the crystal has a highly reflective dielectric coating, so the probe beam makes a double pass of the crystal, doubling the amount of Faraday rotation if the crystal is exposed to a magnetic field with a component along the direction of laser propagation. The probe beam makes a second pass of the beamsplitter and the transmitted component is passed though a second linear polarizer (with zero axis oriented at $45^{\circ}$ to the first) before being transported to a photodiode which is used to measure changes in intensity resulting from Faraday rotation in the crystal. The inset shows the position of the probe crystal relative to the beam in the plane transverse to the beam axis. 
}
\label{fig:Faraday_layout}
\end{figure}


\begin{figure}[h]
\centering
\includegraphics[width=1\textwidth]{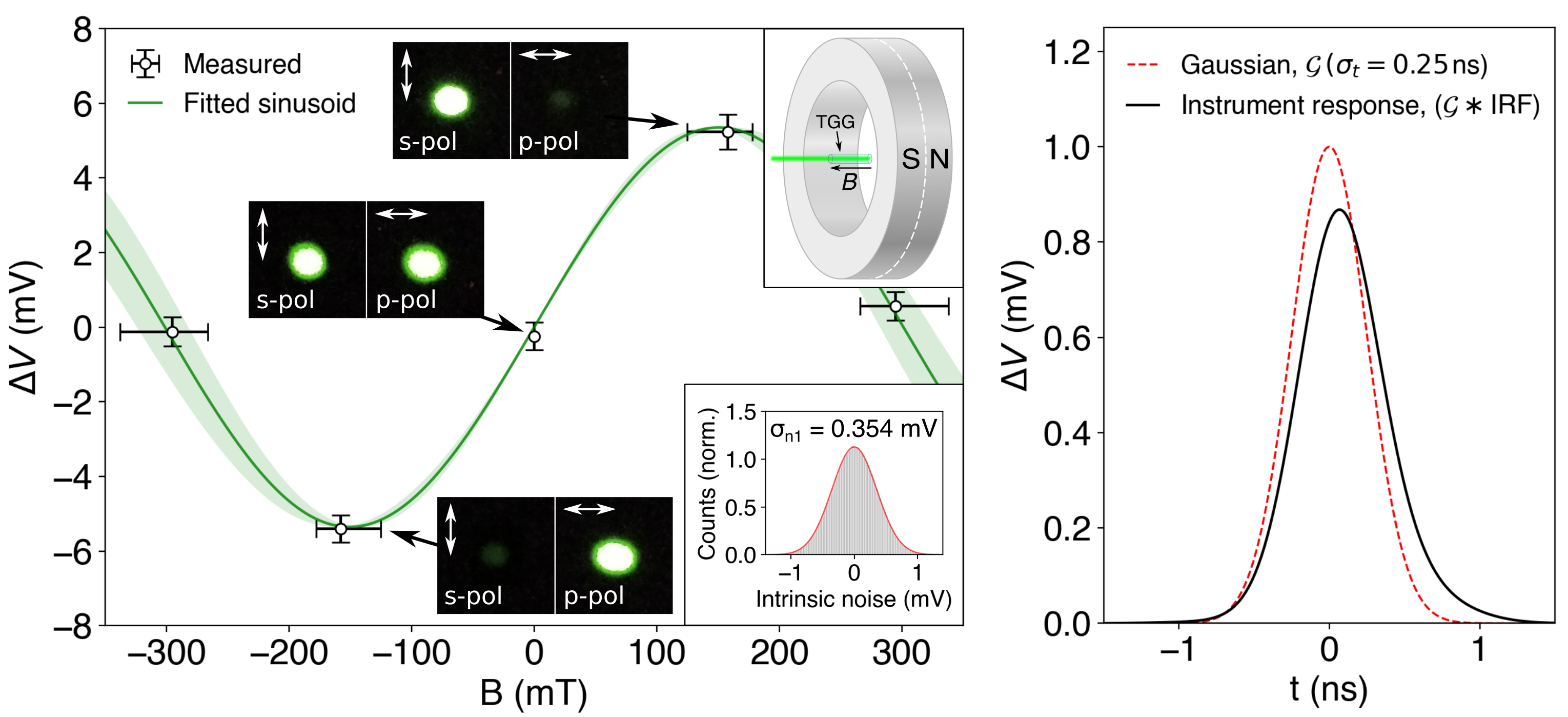}
\caption[Performance of the Faraday probe]{{\bf Performance of the Faraday probe.} Left panel: The Verdet constant of the TGG crystal is measured by fitting a sinusoid to the change in probe beam intensity when exposed to a permanent magnet (as depicted in the upper left inset). The orthogonal components of linear polarization (labelled s-polarized and p-polarized) visibly change in intensity when the crystal is exposed to the magnetic field. The intrinsic electronic noise for a single measurement, which limits the probe sensitivity, is measured to be Gaussian with $\sigma_{\rm n1}=\SI{0.354}{\milli\volt}$, as shown in the lower right inset. Right panel: The instrument response function obtained from impulse measurements with an ultra-fast femtosecond laser is convolved with a Gaussian $\sigma_{t}=\SI{0.25}{\nano\second}$ to show that changes in intensity on timescales of the beam duration can be detected.
}
\label{fig:verdet_performance}
\end{figure}

\begin{figure}[h]
\centering
\includegraphics[width=0.75\textwidth]{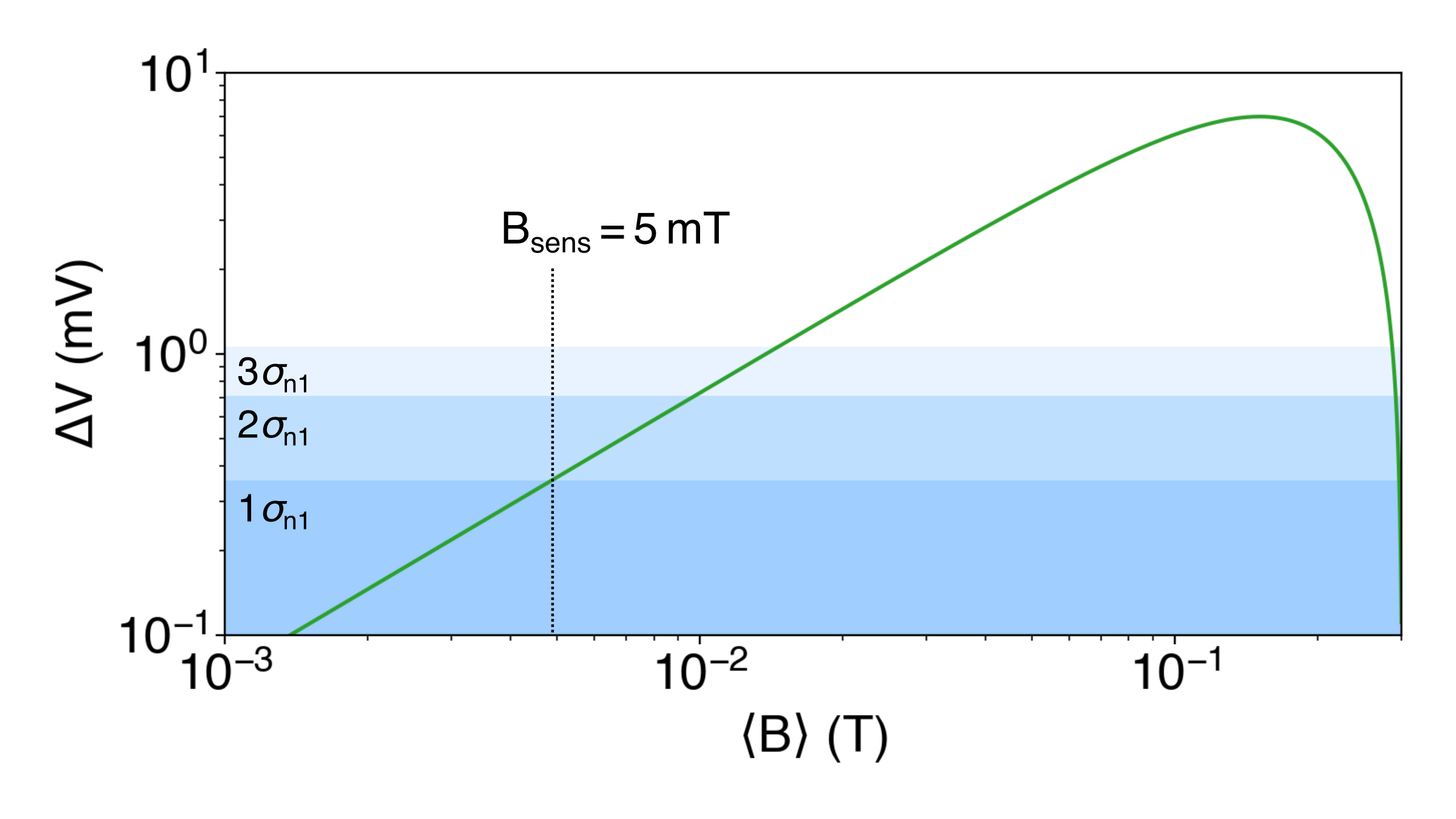}
\caption[Faraday probe sensitivity to magnetic fields]{{\bf Faraday probe sensitivity to magnetic fields.} The change in probe beam intensity ($\Delta V$) is plotted with respect to the mean magnetic field aligned with path of the laser beam through the crystal ($\langle B \rangle$): $\Delta V = V_0\,[\cos^2(\langle B\rangle \mathcal{V}L + 45^{\circ})-\frac{1}{2}]$. The blue-shaded regions correspond to multiples of $\sigma_{\rm n1}$, the intrinsic electronic noise for a single measurement. 
}
\label{fig:Faraday-diagnostic_performance}
\end{figure}

\clearpage

\subsection{Additional Faraday probe and beam profile measurements}
As discussed in the main text, additional magnetic field measurements were obtained with the Faraday probe crystal oriented radially with respect to the beam path (as shown in Supplementary Figure~\ref{fig:data_supp_mag}). Supplementary Figure~\ref{fig:data_supp_screen} shows transverse beam profiles imaged after the plasma when the discharge is operated at different fill pressures to produce different plasma conditions. 

\begin{figure}[h!]
\centering
\includegraphics[width=0.6\textwidth]{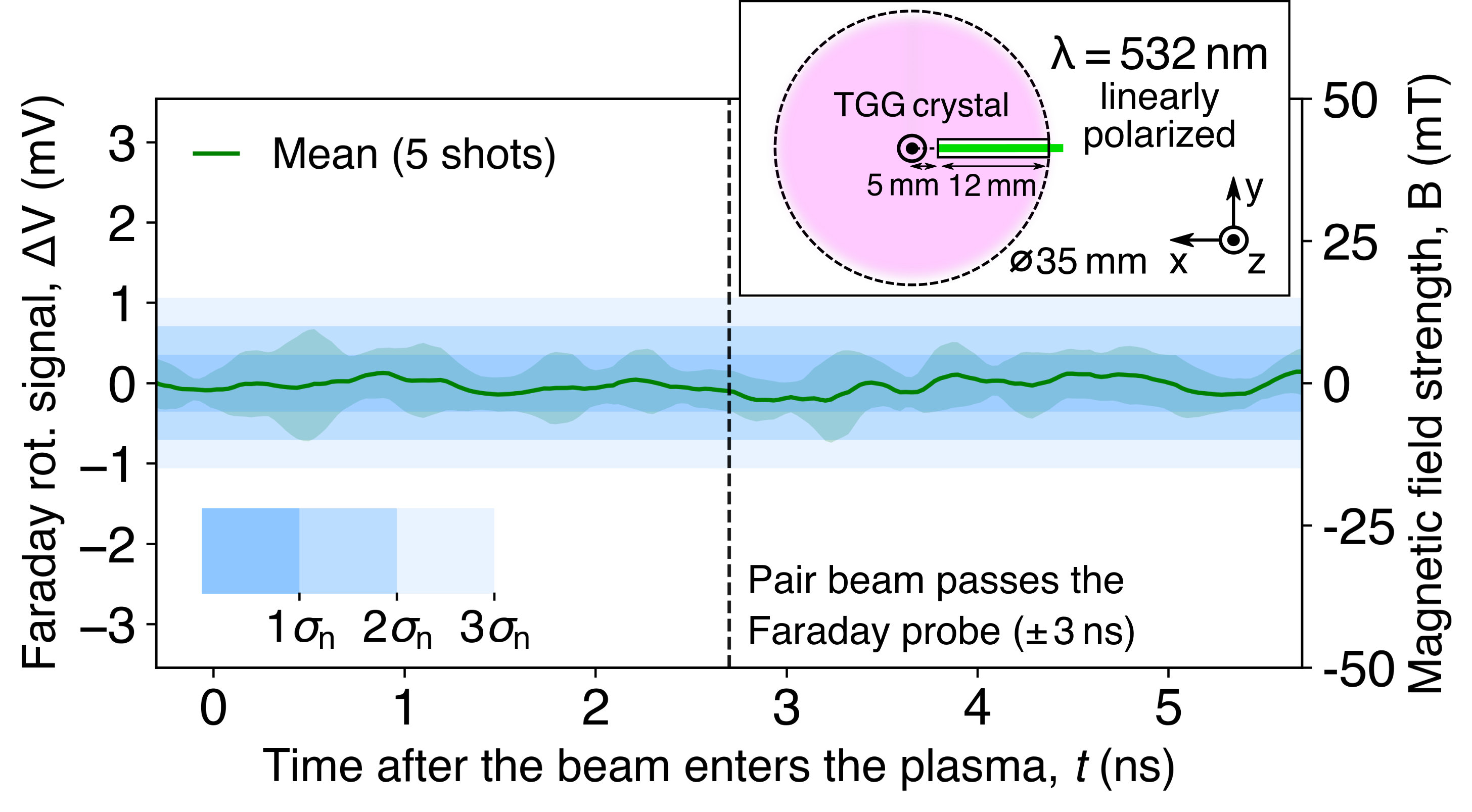}
\caption{{\bf Magnetic field measurements with the Faraday probe oriented radially.} Magnetic fields are measured at the end of the plasma using a time-resolved Faraday rotation technique (see the caption of Figure~\ref{fig:data}). The orientation of the magneto-optic TGG crystal suspended in the plasma is shown in the inset. The mean signal from five shots is plotted, with corresponding standard deviation represented by the green shaded region. The blue shaded regions show the standard deviation of the intrinsic electronic noise.
}
\label{fig:data_supp_mag}
\end{figure}

\begin{figure}[h!]
\centering
\includegraphics[width=0.7\textwidth]{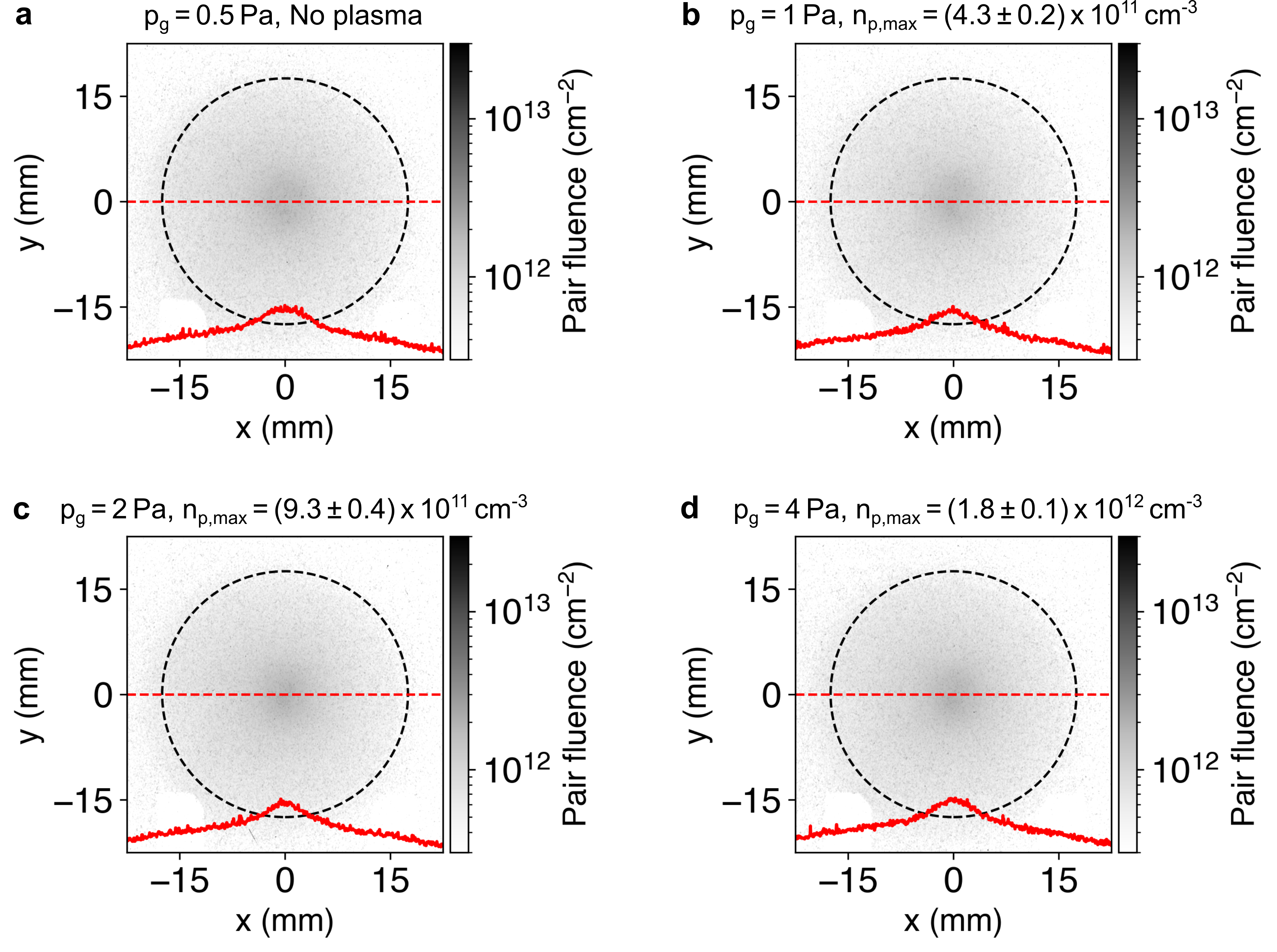}
\caption{{\bf Beam profile measurements using different plasma density conditions.} The transverse beam profile is measured using a Chromox luminescence screen positioned $\SI{90}{\milli\metre}$ downstream of the plasma discharge (see the caption of Figure~\ref{fig:data}). The residual primary proton beam is subtracted from the images, leaving the fluence of electron-positron pairs (plus additional secondaries). Central lineouts are shown (red) for different conditions of the plasma discharge: (a) $p_{\mathrm{g}} = \SI{0.5}{\pascal}$, no plasma, (b) $p_{\mathrm{g}} = \SI{1}{\pascal}$, $n_{\rm p,max}=(4.3\pm0.2)\times\SI{e11}{\per\centi\metre\cubed}$, (c) $p_{\mathrm{g}} = \SI{2}{\pascal}$, $n_{\rm p,max}=(9.3\pm0.4)\times\SI{e11}{\per\centi\metre\cubed}$, and (d) $p_{\mathrm{g}} = \SI{4}{\pascal}$, $n_{\rm p,max}=(1.8\pm0.1)\times\SI{e12}{\per\centi\metre\cubed}$.
}
\label{fig:data_supp_screen}
\end{figure}

\clearpage

\section{Dynamical scaling of the Vlasov-Landau-Maxwell equation}

In the kinetic description of a plasma, the evolution of the distribution function for a species $\alpha$, $f_{\alpha}$, is described by the Vlasov-Landau-Maxwell equation:
\begin{equation}
\dfrac{\partial f_{\alpha }}{\partial t}+\boldsymbol{v} \cdot \nabla f_{\alpha }+q_{\alpha} \left( \boldsymbol{E}+ \boldsymbol{v} \times \boldsymbol{B}\right) \cdot \dfrac{\partial f_{\alpha }}{\partial \boldsymbol{p}} = \bigg(\dfrac{\partial f_{\alpha }}{\partial t}\bigg)_{\rm c},
\label{eq:Vlasov-Landau}
\end{equation}
where $\boldsymbol{E}$ and $\boldsymbol{B}$ are electric and magnetic fields (averaged over microscales) satisfying Maxwell’s equations, $\boldsymbol{p}$ is the particle momentum, $\boldsymbol{v}$ is the particle velocity, and $q_{\alpha}$ is the particle charge. The term $(\partial f_{\alpha }/\partial t)_{\rm c}$ is the collision operator, which is negligibly small for collisionless plasmas provided the free-electron collision frequencies with ions and neutrals are much smaller than the plasma frequency: $\nu_{\rm e}\sim \frac{\lambda_{_{\rm D}}}{\lambda_{\rm mfp}}\omega_{\rm p} \ll \omega_{\rm p}$, where $\lambda_{_{\rm D}}$ is the Debye length and $\lambda_{\rm mfp}$ is the mean free path of electron collisions.
\\
\\
The ratio of the Debye length to the mean free path of electron collisions with ions ($\lambda_{\rm ei}$) and neutrals ($\lambda_{\rm en}$) can be approximated assuming Lennard-Jones potentials for the neutrals:
\begin{equation}
\frac{\lambda_{_{\rm D}}}{\lambda_{\rm ei}} \sim 10^{-12}~\bigg(\frac{T_{\rm e}}{\si{\electronvolt}}\bigg)^{-3/2}\bigg(\frac{n_{\rm e}}{\si{\per\centi\metre\cubed}}\bigg)^{-1/2}\bigg(\frac{n_{\rm i}}{\si{\per\centi\metre\cubed}}\bigg),
\end{equation}
\begin{equation}
\frac{\lambda_{_{\rm D}}}{\lambda_{\rm en}} \sim 10^{-11}~\bigg(\frac{T_{\rm e}}{\si{\electronvolt}}\bigg)^{1/2}\bigg(\frac{n_{\rm e}}{\si{\per\centi\metre\cubed}}\bigg)^{-1/2}\bigg(\frac{n_{\rm n}}{\si{\per\centi\metre\cubed}}\bigg).
\end{equation}
\\
\\
Then we can evaluate the collisionality of the laboratory experiment and the astrophysical case of blazar-induced pair beams in cosmic voids: $\nu_{\rm ei}/\omega_{\rm p}\sim10^{-6}$ and $\nu_{\rm en}/\omega_{\rm p}\sim10^{-3}$ in the experiment, whilst $\nu_{\rm ei}/\omega_{\rm p}\sim10^{-16}$ and $\nu_{\rm en}/\omega_{\rm p}\sim10^{-13}$ in the astrophysical case. In both systems, the collisional terms in the Vlasov-Landau-Maxwell equation can be neglected.

\end{document}